\documentclass[preprint, authoryear]{elsarticle}

\usepackage{natbib}
\usepackage[utf8]{inputenc} 
\usepackage[T1]{fontenc}
\usepackage{lmodern}
\usepackage[fleqn]{mathtools}
\usepackage{rotating}
\usepackage{amssymb}
\usepackage{amsmath}
\usepackage{graphicx}
\usepackage{caption}
\usepackage{epstopdf}
\usepackage{stackrel}
\usepackage{soul}
\usepackage{color}

\newcommand*\diff{\mathop{}\!\mathrm{d}}

\renewcommand*{\underline}{\ul}

\newtheorem{lem}{Lemma}
\newtheorem{cor}{Corollary}
\newdefinition{rmk}{Remark}
\newproof{pf}{Proof}
\newproof{pot}{Proof of Theorem \ref{thm2}}

\begin{document}

\title{Body and Tail \\[12pt]
	{\large Separating the distribution function \\
	by an efficient tail-detecting procedure in risk management}}

\author[hhu]{Ingo Hoffmann\corref{cor1}}
\ead{Ingo.Hoffmann@hhu.de}
\author[hhu]{Christoph J.~B\"orner}
\ead{Christoph.Boerner@hhu.de}

\cortext[cor1]{Corresponding author. Tel.: +49 211 81-15258; Fax.: +49 211 81-15316}

\address[hhu]{Financial Services, Faculty of Business Administration and Economics, \\ Heinrich Heine University D\"usseldorf, 40225 D\"usseldorf,
	Germany}

\begin{abstract}
	In risk management, tail risks are of crucial importance.
	The quality of a tail model, which is determined by data from an unknown distribution, depends critically on the subset of data used to model the tail. Based on a suitably weighted mean square error, we present a method that can separate the required subset. The selected data are used to determine the parameters of the tail model. Notably, no parameter specifications have to be made to apply the proposed procedure. Standard goodness of fit tests allow us to evaluate the quality of the fitted tail model. We apply the method to standard distributions that are usually considered in the finance and insurance industries. In addition, for the MSCI World Index, we use historical data to identify the tail model and to compute the quantiles required for a risk assessment.
\end{abstract}

\begin{keyword}
	Anderson-Darling Statistic \sep Exceedances \sep Extreme Value Theory \sep Generalized Pareto Distribution \sep Quantile Estimation \sep Threshold Selection \sep Risk Management. \\[10pt]
\end{keyword}

\maketitle

\thispagestyle{empty}
\newpage
\section{Introduction}
In many disciplines, there is often a need to adapt a statistical model to existing data to be able to make statements regarding uncertain future outcomes. In particular, when assessing risks, an estimate of major losses must be based on events that, despite having a low probability of occurrence, have a high impact. Since the actual distribution of data -- the parent distribution -- is generally unknown, statisticians begin their modeling with a guess regarding the underlying statistical model. In a first step, they try to fit one or more parametric distribution functions as a model to the data to evaluate the rare events in the next step. These models generally do not perfectly reflect the data. However, specific statistical tests can be applied to assess how well or how poorly a model fits the data as a whole. Nevertheless, especially in the case of rare events and high damage, small uncertainties in the assumption of a model lead to a faulty description of these extremes and will call into question the information value of the approach. Therefore, any uncertainties regarding the underlying model and the resulting misjudgements must be assumed to negatively affect the quality of the statements in many fields of application, especially those interested in models for rare events. Particularly, model uncertainties pose problems in the finance and insurance industries, especially when rare events need to be evaluated, for example, by calculating high quantiles.\\

To make more precise statements regarding rare events and their severity, statisticians can describe the tail of the parent distribution function using a separate model and calculate the corresponding quantiles more precisely to improve the quality of the results. 
In the case of financial institutions, the respective regulatory frameworks provide statisticians and risk managers with the confidence levels of the parent distribution quantiles  \citep{basel04, directive09, directive13, regulation13}. Depending on the purpose, the confidence level is frequently given as 99.9\%; however, the available data often do not cover this area at all. 
The calculation of the capital that is regulatorily required to take a risk is based on the value-at-risk (VaR) or conditional value-at-risk (CVaR), which are calculated from high quantiles. 
In addition to these regulatory risk measures, additional risk measures exist for internal management decisions, such as the return on risk-adjusted capital (RORAC) and the risk-adjusted return on capital (RAROC), which are also calculated from high quantiles of the parent distribution and are important for the risk assessment of a company. Given this framework, a standalone modeling of the tails of the underlying distribution is suggested for more accurate calculation of the risk values.\\

For a very large class of parent distribution functions, the generalized Pareto distribution (GPD) can be used as a model for the tail, cf.,\ e.g.,\ \cite{embrechts03}. This class of distributions includes all common parent distributions that play a role in the financial sector such that almost no uncertainty exists regarding the model selection for the tail of the unknown parent distribution. The required quantiles can then be determined to high confidence levels with sufficient certainty. A certain threshold thus divides the parent distribution into two areas: a body and a tail region. This approach is already common practice for calculating high quantiles more accurately, as indicated in \citet{basel09}. Below the threshold, evaluations are performed based on a statistical model assumed for the body of the unknown parent distribution; above the threshold, analyses are performed using the GPD as a model for the tail.\\

In practice, for the separate modeling of the tail, only the parameters of the GPD need to be determined from the data pertaining to the tail region of the parent distribution. This can be carried out using, e.g.,\ the maximum likelihood method. However, a pivotal question remains: which of the available data belongs to the tail of the parent distribution, and which, to the body? In other words, at what threshold $u$ does the data belong to the tail of the parent distribution? The answer to this question is crucial in terms of the quality and validity of the model because the shape parameter $\xi$ of the GPD (sometimes called the tail parameter) depends strongly on the correct threshold being chosen. The threshold parameter decisively determines the values of the high quantiles. Estimation errors regarding this parameter result in significant estimation errors in the quantiles and thus lead to significant errors in the calculation of the capital required to take the corresponding risk.
Estimation errors occur both when the set of data points belonging to the tail of the parent distribution chosen is too small and when the set is too large. In the first case, statistical errors and small sample effects greatly reduce the accuracy, while in the second case, the GPD may no longer be the correct model for the tail of the parent distribution.\\

The need for a suitable and efficient method for determining the optimum threshold $u$ is emphasized by many practitioners. However, they also note that no definitive best practice currently exists \citep{mcneil97, embrechts03, defontnouvelle05, chernobai06, dutta07}.
Of course, some approaches have already been developed to address the problem of choosing the optimal threshold. \citet{pickands75} considers absolute distances between the ``empirical upper tail`` distribution function and the GPD and suggests that the threshold should be chosen such that the distance becomes minimal. For an unknown threshold, \citet{hill75} suggests a successive hypothesis test using a standard goodness of fit method. The idea is to increase the amount of data for the tail until the goodness of fit test rejects the hypothesis that the GPD is the model for the tail of the unknown parent distribution. A similar approach is proposed in the work of \citet{choulakian01}. \citet{smith87} exploits the property of the maximum likelihood estimator of the shape parameter and can implicitly deduce the possible threshold value using his method. 
In other fields of research, the characteristic of the GPD mean excess function is exploited to determine the optimal threshold. From a sufficiently high threshold value, the mean excess function should be linear and thus mark the optimum threshold value. However, statistical influences and small sample effects terminate the linear behavior, and the optimal threshold can be determined only with great uncertainty; see, e.g., \citet{embrechts03} and the references cited therein.
\citet{danielsson01} proposes a method based on the Hill estimator \citep{hill75} for the shape parameter of the GPD. In a sorted data series, the mean square error is evaluated as the subset of data allocated to the tail increases. The optimum threshold is found when the mean square error is at its minimum. 
\citet{nguyen12} uses statistical test methods to evaluate the property wherein, starting at a certain threshold, the sorted data belonging to the tail of the distribution statistically behave like points of a Poisson random measure with a power intensity. This property can then be statistically tested as the subset increases in size and is aborted as soon as the test rejects the hypothetical property. 
As far as can be seen, none of the existing procedures have prevailed \citep{dutta07}.
Therefore, a simple, efficient and always applicable method for determining the threshold value is still needed.\\

The aim of this study is to develop a statistically based, efficient method that optimally determines the required threshold from the available, measured data. The starting points of our paper are some of the already developed procedures, which we want to combine and extend appropriately. Methods based on the minimization of a distance measure or on statistical tests seem to be most promising. Consequently, we propose a combined method for determining the threshold, which includes the minimum of a distance measure and a suitable statistical test. As well as the use of statistical tests in the framework of goodness of fit procedures \citep{vonmises31, anderson54, shorack09}, the determination of distribution parameters via minimum distance methods \citep{wolfowitz57, blyth70, parr80, boos82} from the field of decision theory \citep{ferguson67} is well established.
Both research areas are essentially based on the weighted mean square error, which evaluates the distance between an empirical distribution function of the data and a suspected parent distribution. Whereas in the first case, the distribution of the weighted mean square error leads to the definition of critical values and thus to the well-known goodness of fit tests, e.g.,\ the Cram\'er-von Mises test \citep{cramer28, vonmises31} or Anderson-Darling test \citep{anderson52, anderson54}, in the second case, the parameters of the assumed parent distribution are determined by minimizing the error \citep{boos82}.\\

In our paper, we select a special weight function for the squared errors, which weights deviations in the tail of the parent distribution more heavily. With this specially chosen function, the weighted mean square error is a measure of the total deviation between the empirical distribution and the fitted GPD. We evaluate this total deviation as the data size in the ordered sample increases. The minimum of this total deviation then marks the optimal threshold. Once the threshold has been determined, the GPD is adapted
as a model for the tail on the data associated with the tail region. In addition, we perform statistical tests to evaluate the quality of the modeling.\\

The remainder of the paper is structured as follows:
In the first sections (from \ref{TestStatistics} to \ref{CV}), we develop the theoretical foundations for the tail-detection method, which we propose based on the existing theories of various disciplines.

Section \ref{TestStatistics} shows how the weighted mean square error is explicitly computed when depending on a weight function. The weight function is specially chosen for our purposes and is equipped with a free positive stress parameter. This stress parameter allows, at a certain position on the distribution function, changing the strength with which the deviations from the empirical distribution function at that position are weighted. 
Put simply,
using the stress parameter, the magnification is adjusted, with which the deviation between the distributions at a fixed position is considered. If we initially leave the stress parameter indeterminate in the calculation of the weighted mean square error, two families of statistics can be derived. One family of statistics is for the lower tail; the other, for the upper tail. As a theoretical interim result, we show that these two families can be transformed into each other via coordinate transformation.
Section \ref{RF} uses decision theory methods to find the most suitable stress parameter for our purposes. From the families of statistics, the statistic that belongs to this stress parameter is used to determine the threshold based on the available data. In addition, the relationship with the standard goodness of fit tests is shown, which we use to evaluate the final result.
Section \ref{GPD} briefly summarizes the key features of the GPD.
Since we expect small values for the weighted mean square error, which will be reflected in the standard goodness of fit test, we need critical values at high confidence levels to perform the tests. These are currently not tabulated; thus, in Section \ref{CV} we determine the required critical values.\\

After presenting all theoretical bases in the first sections and creating the necessary technical prerequisites, the subsequent sections introduce the method of tail detection and demonstrate its application via examples.\\

Section \ref{TD} presents the procedure for determining the threshold value and the tail model in detail. Applications are shown for different parent distributions. Here, we focus on parent distributions that are commonly used in the finance and insurance industries. The properties of the method are investigated via Monte Carlo simulations.
Section \ref{examples} shows the application of the method to single-row data. 
As a practical example with real data, the MSCI World Index is considered.
We consider the distribution of the MSCI World Index as unknown. In this example we do not fit a parametric model to estimate the parent distribution but focus directly on modeling the tail.
Based on the historical data, the tail model is determined using the previously developed method. The VaR and CVaR are then determined using the quantiles at high confidence levels.
The last section discusses the results and summarizes the key points.
\section{Definition of test statistics}\label{TestStatistics}
Due to regulatory requirements, the financial industry is interested in finding a statistical model for the data collected by risk management that has high quality, especially at high quantiles. This requires statistical tests that take special account of deviations in the upper or lower tail. In this section, we derive the two corresponding families of test statistics. Furthermore, these two statistics are shown to have symmetry and to be very well suited for detecting the beginning of tails.
\subsection{Empirical distribution function}\label{EDF}
Let $X_1, X_2, \ldots , X_n$ be a sample of random variables with common unknown continuous distribution function $F(x)$ and density function $f(x)$. The corresponding empirical distribution function for $n$ observations is defined as
\begin{flalign}\label{EDFEQ}
F_n(x)& = \frac 1 n \sum_{i = 1}^n {\boldsymbol 1} (X_i \leq x),
\end{flalign}
where ${\boldsymbol 1}$ is the indicator function, with ${\boldsymbol 1} (X_i \leq x)$ equal to one if $X_i\leq x$ and zero otherwise. Thus, $F_n(x) = \frac k n$ if $k$ observations are lower than or equal to $x$ for $k=0,1, \ldots ,n$ \citep{kolmogorov33}.
\subsection{Weighted mean square error}\label{WMSE}
As a convenient measure of the discrepancy or ``distance`` between the two distribution functions $F_n(x)$ and $F(x)$, we consider the weighted mean square error
\begin{flalign}\label{WMSEEQ}
\hat R_n & = n \int_{-\infty}^{+\infty} \left(F_n(x)-F(x) \right)^2\; w(F(x))\;\diff F(x),
\end{flalign}
introduced in the context of statistical test procedures by \citet{cramer28}, \citet{vonmises31} and \citet{smirnov36}. 
The non-negative weight function $w(t)$ in Eq.\ (\ref{WMSEEQ}) is a suitable preassigned function for accentuating the difference between the distribution functions in the range where the test procedure is desired to have sensitivity. Consider the weight function
\begin{flalign}\label{WFEQ}
w(t) & = \frac{1}{t^a(1-t)^b}
\end{flalign}
for real-valued stress parameters $a,b\geq 0$ and $t\in[0,1]$. Here, $a$ affects the weight at the lower tail, and $b$, at the upper tail. Then, for $a=b=0$, Eq.\ (\ref{WMSEEQ}) provides the Cram\'er-von Mises statistic \citep{cramer28, vonmises31}, while when heavily weighting the tails ($a=b=1$), it is equal to the Anderson-Darling statistic \citep{anderson52, anderson54}. The Anderson-Darling statistic weights the difference between the two distributions simultaneously more heavily at both ends of the distribution $F(x)$.\\

Mixed weight functions can hinder the individual study of either one or the other tail of the distribution function. When determining the start of a tail, pure weight functions that focus on one side of the distribution function are beneficial. Therefore, to determine the beginning of the tails of $F(x)$, we use statistics that weight the deviations at either the upper or the lower tail more heavily. The following weight functions are desired and are thus further investigated.\\

Weight function for the lower tail ($a\geq 0, b=0$):
\begin{flalign}\label{WFLTEQ}
w(t) & = \frac{1}{t^a} 
\end{flalign}

Weight function for the upper tail ($a=0, b\geq 0$):
\begin{flalign}\label{WFUTEQ}
w(t) & = \frac{1}{(1-t)^b} 
\end{flalign}

\subsection{Lower tail statistics}\label{LTST}
With Eq.\ (\ref{WFLTEQ}), the weighted mean square error Eq.\ (\ref{WMSEEQ}) reduces to
\begin{flalign}\label{LTWMSEEQ}
\hat R_{n,a,0} & = n \int_{-\infty}^{+\infty} \frac{\left(F_n(x)-F(x) \right)^2}{\left(F(x)\right)^a}\; \diff F(x).
\end{flalign}
Computing formulae for this family of lower tail statistics can be obtained by following the method given in \citet{anderson54}. \\

Let $x_{(1)} \leq x_{(2)}\leq \ldots \leq x_{(n)}$ be the sample values (in ascending order) obtained by ordering each realization $x_1, x_2, \ldots , x_n$ of $X_1, X_2, \ldots , X_n$. Then, we can summarize the following calculation rules for the statistics:\\
\\
$\bullet$  {$a\neq 1,2,3$}%
\begin{flalign}\label{LTSEEQ}
\hat R_{n,a,0} & = \frac{2}{(1-a)(2-a)(3-a)}\; n \\\nonumber
& + \sum_{i=1}^n\left[ \frac{2}{2-a}\left(F(x_{(i)})\right)^{2-a} - \frac{2i-1}{n}\frac{1}{1-a} \left(F(x_{(i)})\right)^{1-a} \right]
\end{flalign}
Note: In the special case where $a=0$, Eq.\ (\ref{LTSEEQ}) reduces to the statistics $W_n^2\; (= \hat R_{n,0,0})$ proposed by \citet{cramer28} and \citet{vonmises31}:
\begin{flalign}\label{CVM}
W_n^2  & = \frac{1}{12n}  
+ \sum_{i=1}^n\left[ \frac{2i-1}{2n} - F(x_{(i)})  \right]^2
\end{flalign}
$\bullet$ {$a= 1$}%
\begin{flalign}\label{LTSEEQa1}
\hat R_{n,1,0} & = -\frac{3}{2} n 
 + \sum_{i=1}^n\left[ 2 F(x_{(i)}) - \frac{2i-1}{n} \ln\left(F(x_{(i)})\right) \right]
\end{flalign}
For the purpose of obtaining an appropriate goodness of fit test specifically for the tail of a distribution, the computation formulae Eq.\ (\ref{LTSEEQa1}) and Eq.\ (\ref{LTSEEQb1}) were first described by \citet{ahmad88} and later examined more formally by the same authors with regard to the distribution of their test statistics $AL_n^2 \; (= \hat R_{n,1,0})$ and $AU_n^2\; (= \hat R_{n,0,1})$, respectively \citep*{sinclair90}.\\
\\
$\bullet$ {$a=2$}%
\begin{flalign}\label{LTSEEQa2}
\hat R_{n,2,0} & = 
\sum_{i=1}^n\left[\frac{2i-1}{n}\frac{1}{F(x_{(i)})}
+ 2 \ln\left(F(x_{(i)})\right) \right]
\end{flalign}
$\bullet$ {$a=3$}\\
\\
For the stress parameter $a = 3$, no feasible solution can be calculated because $\hat R_{n,3,0}$ approaches infinity.
\subsection{Upper tail statistics}\label{UTST}
With Eq.\ (\ref{WFUTEQ}), the weighted mean square error Eq.\ (\ref{WMSEEQ}) becomes
\begin{flalign}\label{UTWMSEEQ}
\hat R_{n,0,b} & = n \int_{-\infty}^{+\infty} \frac{\left(F_n(x)-F(x) \right)^2}{\left( 1 - F(x)\right)^b}\;\diff F(x).
\end{flalign}
As with Eq.\ (\ref{LTWMSEEQ}), the statistics for the upper tail can now be calculated using Eq.\ (\ref{UTWMSEEQ}).\\
\\
$\bullet$ {$b\neq 1,2,3$}%
\begin{flalign}\label{UTSEEQ}
\hat R_{n,0,b} & = \frac{2}{(1-b)(2-b)(3-b)}\; n \\\nonumber
& + \sum_{i=1}^n\left[ \frac{2}{2-b}\left(1-F(x_{(i)})\right)^{2-b} - \frac{2(n-i)+1}{n}\frac{1}{1-b} \left(1-F(x_{(i)})\right)^{1-b} \right]
\end{flalign}
Note: When $b = 0$, then Eq.\ (\ref{UTSEEQ}), like Eq.\ (\ref{LTSEEQ}), reduces to the known statistic $W_n^2$ \citep{cramer28, vonmises31}, cf.\ Eq.\ (\ref{CVM}).\\
\\
$\bullet$ {$b= 1$}%
\begin{flalign}\label{LTSEEQb1}
\hat R_{n,0,1} & = \frac{1}{2} n 
- \sum_{i=1}^n\left[ 2 F(x_{(i)}) + \frac{2(n-i)+1}{n} \ln\left(1-F(x_{(i)})\right) \right]
\end{flalign}
Because of the identity $w(t) = \frac{1}{t(1-t)}= \frac{1}{t} + \frac{1}{1-t}$, the expression $\hat R_{n,1,1} = \hat R_{n,1,0} + \hat R_{n,0,1} $ reduces, with appropriate renumbering, to the well-known Anderson-Darling statistic $A_n^2 \; (= \hat R_{n,1,1})$  \citep{anderson52, anderson54}:
\begin{flalign}\label{ADTS}
A_{n}^2 & = -n 
- \sum_{i=1}^n \frac{2i-1}{n} 
\left[  \ln\left(F(x_{(i)})\right) 
      + \ln\left(1-F(x_{(n-i+1)})\right) \right].
\end{flalign}
$\bullet$ {$b=2$}%
\begin{flalign}\label{LTSEEQb2}
\hat R_{n,0,2} & = 
\sum_{i=1}^n\left[\frac{2(n-i)+1}{n}\frac{1}{1-F(x_{(i)})}
+ 2 \ln\left(1-F(x_{(i)})\right) \right]
\end{flalign}
$\bullet$ {$b=3$}\\
\\
As before, for the stress parameter $b = 3$, no feasible solution can be calculated because $\hat R_{n,0,3}$ approaches infinity.\\

The families of upper and lower tail statistics defined so far are based on the proper choice of a weight function $w$. 
To emphasize the deviation error in the tails more strongly, special weight functions are used, as illustrated by Eq.\ (\ref{WFLTEQ}) and Eq.\ (\ref{WFUTEQ}). As the following lemma shows, these weight functions have internal symmetry that translates to the weighted mean square error and that can be exploited in practice.
\begin{lem}\label{lemmasym}
  Let $\hat R_{X;n,a,0}$ be the weighted mean square error defined by Eq.\ (\ref{LTWMSEEQ}) for random variable $X$, which comes from a continuous distribution function $F_X(x)$. Furthermore, let $Z = -X$ be a coordinate transformation of the random variable. Then, $\forall a \in\mathbb{R}$,
  \begin{flalign}\label{WMSESYM}
  \hat R_{X;n,a,0} = \hat R_{Z;n,0,a}.
  \end{flalign}
\end{lem}
\begin{pf}
The coordinate transformation leads to a substitution of the variables $X$ in the integrals. First, substitution of the continuous and empirical distribution function is performed, resulting in the substitution rules used in Eq.\ (\ref{LTWMSEEQ}) (or Eq.\ (\ref{UTWMSEEQ})): $F_X(x)\rightarrow 1 - F_Z(z)$, $\diff F_X(x)\rightarrow -\diff F_Z(z)$ and $F_n(x)\rightarrow 1 - F_n(z)$. Finally, the domain of integration is changed according to the coordinate transformation, and because of the minus sign, the orientation of the integration domain is reversed. This results in the representation of the weighted mean square error as in Eq.\ (\ref{UTWMSEEQ}) (or Eq.\ (\ref{LTWMSEEQ})) for the random variable $Z$. $\blacksquare$
\end{pf}
The practical advantage of lemma \ref{lemmasym} is that only one family of statistics -- either for the upper tail or the lower tail -- requires further investigation. Furthermore, lower tail analyses of a distribution of the random variables $X$, for example, can be performed by using one of the upper tail statistics for the random variable $Z$, after utilizing the transformation $Z = -X$.
Therefore, only one of the statistical families needs to be considered during software implementation.

\begin{cor}
	The Cram\'er-von Mises statistic $W_{X;n}^2$ and the Anderson-Darling statistic $A_{X;n}^2$ are invariant and remain unchanged when the transformation $Z=-X$ is applied to the random variable $X$.
\end{cor}	
\begin{pf}
For the Cram\'er-von Mises statistic, $W_{X;n}^2 = \hat R_{X;n,0,0} $ holds. According to lemma \ref{lemmasym}, the transformation $Z=-X$ leads to $\hat R_{Z;n,0,0}$, which is equal to $W_{Z;n}^2$. No change in the last two indices occurs; thus, the computing formula Eq.\ (\ref{CVM}) does not change. In the case of the Anderson-Darling statistic, $A_{X;n}^2 = \hat R_{X;n,1,1} = \hat R_{X;n,1,0} + \hat R_{X;n,0,1}$ holds. The transformation then results in $\hat R_{Z;n,0,1} + \hat R_{Z;n,1,0} = \hat R_{Z;n,1,1}$, which is equal to $A_{Z;n}^2$. No change in the last two indices occurs; thus, the computing formula Eq.\ (\ref{ADTS}) does not change.
$\blacksquare$
\end{pf}	
In general, the two families of statistics are not invariant under a transformation $Z = -X$, i.e.,\ they change their appearance. Due to lemma \ref{lemmasym}, for a pregiven stress parameter $a$, the statistic for the lower tail of $F_X(x)$ changes into the statistic for the upper tail of $F_Z(z)$. The following corollary provides the necessary transformations in the corresponding computation formulae.
\begin{cor}
Let the stress parameter $a$ be fixed, and apply $Z = -X$ to the random variable $X$. Then, the computation formulae for the two families of statistics can be transformed into each other by using the following substitutions: change the summation index from $i$ to $k$, and then set
$F_X(x_{(k)})\rightarrow 1 - F_Z(z_{(k)})$. The single number $i$, which results from the numerator of the empirical distribution function, must be changed according to $i\rightarrow n-k+1 $.
\end{cor}
\begin{pf}
The substitution $F_X(x_{(k)})\rightarrow 1 - F_Z(z_{(k)})$ follows directly from the proof of lemma \ref{lemmasym}. Furthermore, the transformation $Z=-X$ on the sample of random variables $Z_i= - X_i$ leads to a new order statistic $Z_{(i)}$, and a realization $-x_{(i)}$ appears at a new position $z_{(k)} = -x_{(i)}$. Then, the following identity holds: 
\begin{flalign*}
\begin{matrix*}[l]
1 &=& \frac 1 n \displaystyle\sum_{j = 1}^n {\boldsymbol 1} (Z_{j} \geq z_{(k)}) 
  &+& \frac 1 n \displaystyle\sum_{j = 1}^n {\boldsymbol 1} (Z_{j} < z_{(k)}) \\[12pt]
  &=& \frac 1 n \displaystyle\sum_{j = 1}^n {\boldsymbol 1} (-z_{(k)} \geq -Z_{j})   &+& \frac{k-1}{n} \\[12pt]
  &=& \frac 1 n \displaystyle\sum_{j = 1}^n {\boldsymbol 1} (x_{(i)} \geq X_{j})   &+& \frac{k-1}{n}  \\[12pt]
  &=& \frac{i}{n}   
  &+&  \frac{k-1}{n}.
\end{matrix*}
\end{flalign*}
The last equation leads to the desired transformation of the position index $i = n-k+1$. By simple algebraic transformations, with the given substitutions for a fixed stress parameter $a$, the corresponding computation formulae of the upper and lower tail statistic can be transformed into one another.
$\blacksquare$
\end{pf}
With these preliminary considerations, it is sufficient to carry out further investigations on only one of the two statistical families.
\section{Selection of the appropriate stress parameter}\label{RF}
In this section, we explore the following question: which stress parameter should be chosen, from a theoretical and practical point of view, when addressing financial and insurance issues? 
Since we focus on the determination of the beginning of the tail region of a distribution for small values of a selected statistic, a suitable method for assessing the quality of a stress parameter may involve using the so-called risk function $R$ \citep{aggarwal55}. The risk function is a tried and tested method in decision theory for determining the average loss when statisticians set their model for given data \citep{ferguson67}. In this school of thought within decision theory, the weighted mean square error $\hat R_n$ defined in the previous section (see Eq.\ (\ref{WMSEEQ})) is generally referred to as the loss function, and the expected value of the loss function is called the risk function:
\begin{flalign}\label{RFRn}
R_n & = {\text E} \left[ \hat R_n  \right].
\end{flalign}
For the case considered here, the risk function can be calculated explicitly. The result summarizes the following lemma and the complementary corollaries.
\begin{lem}\label{lemmariskfunction}
	Let $\hat R_{n,a,0}$ be the weighted mean square error defined by Eq.\ (\ref{LTWMSEEQ}). Then, $\forall a\in \mathbb{R}^{\geq 0}$, the risk function is given by
	\begin{flalign}\label{RFLT}
	R_{n,a,0} = \frac{1}{(2-a)(3-a)}.
	\end{flalign}
\end{lem}
\begin{pf}
	Using the transformation $u=F(x)$, the lower tail statistics can be expressed in terms of $u \in [0, 1]$, and $u_{(1)} \leq u_{(2)}\leq \ldots \leq u_{(n)}$ is an ordered sample of size $n$ from a continuous uniform distribution over the interval $[0, 1]$. The expectation in Eq.\ (\ref{RFRn}) has to be taken with respect to this distribution. Since the distribution of the $i$th order statistic $U_{(i)}$ in a random sample of size $n$ from the uniform distribution over the interval $[0, 1]$ is a beta distribution with probability density 
	\begin{flalign}\label{OSPD}
	p(u) = \frac{1}{B(i, n-i+1)} u^{i-1}(1-u)^{n-i},
	\end{flalign}
	the expectation value for $\hat R_{n,a,0}$ can be calculated as follows:\\
	\\
	$\bullet$ {$a\neq 1,2,3$}\\
	\begin{flalign}\label{lem2Profa}
	R_{n,a,0} & = {\text E} \left[ \hat R_{n,a,0}\right] \\ \nonumber
		      & = \frac{2n}{(1-a)(2-a)(3-a)}\;  \\ \nonumber
		      & \qquad + \sum_{i=1}^n \frac{2}{2-a}{\text E} \left[u_{(i)}^{2-a}\right] \\ \nonumber
		      & \qquad - \sum_{i=1}^n \frac{2i-1}{n}\frac{1}{1-a} {\text E} \left[u_{(i)}^{1-a}\right] \\ \nonumber
		      & = \frac{2n}{(1-a)(2-a)(3-a)}\;  \\ \nonumber
		      & \qquad + \sum_{i=1}^n \frac{2}{2-a}\frac{\int_0^1 u^{i+1-a}(1-u)^{n-i} \diff u }{B(i, n-i+1)} \\ \nonumber
		      & \qquad - \sum_{i=1}^n \frac{2i-1}{n}\frac{1}{1-a} \frac{\int_0^1 u^{i-a}(1-u)^{n-i} \diff u }{B(i, n-i+1)} \\ \nonumber
		      & = \frac{2n}{(1-a)(2-a)(3-a)}\;  \\ \nonumber
		      & \qquad + \sum_{i=1}^n \frac{2}{2-a}\frac{B(i+2-a, n-i+1) }{B(i, n-i+1)} \\ \nonumber
		      & \qquad - \sum_{i=1}^n \frac{2i-1}{n}\frac{1}{1-a} \frac{B(i+1-a, n-i+1)}{B(i, n-i+1)}  \nonumber
	\end{flalign}
	The remaining sums over the quotients of beta functions can be explicitly calculated when the beta functions are expressed in terms of the gamma function \citep{abramowitz14}. The poles must be considered, and the gamma function should be considered in its analytic continuation. If the representation of the gamma functions in terms of the Pochhammer symbol for the rising factorials is used here, the sums can be suitably changed via algebraic transformations. Hence, only the Chu-Vandermonde theorem has to be applied to calculate the result of the sums \citep[Ch.\ 18]{oldham09}. After the outcomes of the sums have been summarized in toto, the asserted relationship Eq.\ (\ref{RFLT}) is obtained.  \\
	\\
	$\bullet$ {$a=1$}\\
	\begin{flalign}\label{lem2Profa1}
	R_{n,1,0} & = {\text E} \left[ \hat R_{n,1,0}\right] \\ \nonumber
	& = -\frac{3n}{2} + \sum_{i=1}^n 2\;{\text E} \left[u_{(i)}\right] 
	  - \sum_{i=1}^n \frac{2i-1}{n}\; {\text E} \left[\ln{u_{(i)}}\right] \\ \nonumber
	& = -\frac{3n}{2} + \sum_{i=1}^n 2\frac{B(i+1, n-i+1)}{B(i, n-i+1)} 
	  - \sum_{i=1}^n \frac{2i-1}{n}  \left( \psi(i) - \psi(n+1) \right)
	\end{flalign}
	For the last sum, we use the computation formulae presented by 
	\citet[Eq.\ (59) therein]{aggarwal55}, with $\psi(i)$ being the digamma function, cf.\ \citet{abramowitz14}. The remaining sums can then be further simplified as follows:
	\begin{flalign}\label{lem2Profa12}
	R_{n,1,0} & = -\frac{3n}{2} + n   +\frac{n}{2} +\frac{1}{2} \\ \nonumber
	& = \frac{1}{2}
	\end{flalign}
	This result is also yielded by Eq.\ (\ref{RFLT}) for $a = 1$.\\
	\\
	$\bullet$ {$a=2$}\\
	\begin{flalign}\label{lem2Profa2}
	R_{n,2,0} & = {\text E} \left[ \hat R_{n,2,0}\right] \\ \nonumber
	& = \sum_{i=1}^n \frac{2i-1}{n}{\text E} \left[\frac{1}{u_{(i)}}\right] 
		+ \sum_{i=1}^n 2\; {\text E} \left[\ln{u_{(i)}}\right] \\ \nonumber
	& = \sum_{i=1}^n \frac{2i-1}{n} \frac{B(i-1, n-i+1)}{B(i, n-i+1)} 
		+ \sum_{i=1}^n 2\; \left( \psi(i) - \psi(n+1) \right) \\ \nonumber
	& = \sum_{i=1}^n \frac{2i-1}{n} \frac{n}{i-1}  -2n
	\end{flalign}
	The last expression shows that the sum becomes infinite, as the first term for $i = 1$ represents a pole, while all other terms of the sum remain finite. This result is described by the pole in Eq.\ (\ref{RFLT}) when $a = 2$.\\
	\\
	$\bullet$ {$a=3$}\\
	\\
	In Section \ref{LTST}, we showed that $\hat R_{n,3,0}$ approaches infinity. The same applies to the expected value $R_{n,3,0}$. This result is described by the pole in Eq.\ (\ref{RFLT}) when $a = 3$.
	$\blacksquare$
\end{pf}
Because of the symmetry of the two families of statistics from Section \ref{TestStatistics}, we note the following:
\begin{cor}\label{RFUTCorrolar}
    Let $\hat R_{n,0,b}$ be the weighted mean square error defined by Eq.\ (\ref{UTWMSEEQ}). Then, $\forall b\in \mathbb{R}^{\geq 0}$, the risk function is given by
	\begin{flalign}\label{RFUT}
		R_{n,0,b} = \frac{1}{(2-b)(3-b)}.
	\end{flalign}
\end{cor}
\begin{pf}
	\begin{flalign}\label{ProofRFUT}
	R_{n,0,b} & = {\text E} \left[ \hat R_{n,0,b}\right] \stackrel{\text{(lemma \ref{lemmasym})}}{=} {\text E} \left[ \hat R_{n,b,0}\right] \stackrel{\text{(lemma \ref{lemmariskfunction})}}{=} \frac{1}{(2-b)(3-b)} \;\;\; \blacksquare
	\end{flalign}	
\end{pf}
In the special cases of the Cram\'er-von Mises statistic and the Anderson-Darling statistic, the following holds:
\begin{cor}\label{CVMCorrolar}
	Let $W_n^2=\hat R_{n,0,0}$ be the Cram\'er-von Mises statistic Eq.\ (\ref{CVM}). Then, the risk function is given by
	\begin{flalign}\label{CVMRF}
	R_{n,0,0} = \frac{1}{6}.
	\end{flalign}
\end{cor}
\begin{pf}
	By inserting $a = 0$ in Eq.\ (\ref{RFLT}). (Note: The risk function calculated here is in accordance with the result of \citet{aggarwal55}.) $\blacksquare$
\end{pf}
\begin{cor}\label{ADCorrolar}
	Let $A_n^2=\hat R_{n,1,1}$ be the Anderson-Darling statistic Eq.\ (\ref{ADTS}). Then, the risk function is given by
	\begin{flalign}\label{ADRF}
	R_{n,1,1} = 1.
	\end{flalign}
\end{cor}
\begin{pf}
	\begin{flalign}\label{ProofRFAD}
	R_{n,1,1} & = {\text E} \left[ \hat R_{n,1,1}\right] = 
				  {\text E} \left[ \hat R_{n,1,0} + \hat R_{n,0,1}\right] =
				  {\text E} \left[ \hat R_{n,1,0}\right] + {\text E} \left[ \hat R_{n,0,1}\right] 
	\end{flalign}
	The last expression leads to
	$R_{n,1,1}=\frac{1}{2} + \frac{1}{2} = 1$, in accordance with the result of \citet{aggarwal55}. $\blacksquare$
\end{pf}

\begin{figure}[htbp]
	\centering
	\captionsetup{labelfont = bf}
	\includegraphics[width=0.975\textwidth]{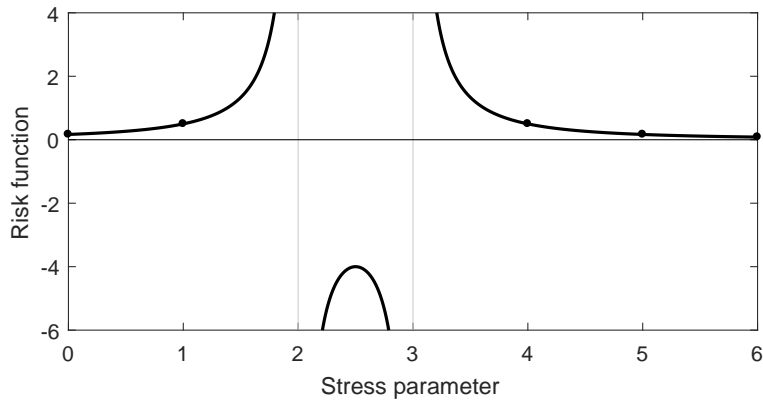}
	\begin{quote}
		\caption[Sample-Size 100]{\label{riskfunctionbild} The dependence of the risk function on the stress parameter (for $a$ or $b$ alike). For integer values, the corresponding risk values are indicated by the bullets. The two poles are denoted by the thin lines.}
	\end{quote}
\end{figure}

To summarize the results, Fig.\ \ref{riskfunctionbild} shows the dependence of the risk function on the stress parameter.
The risk function is symmetric about the local maximum at $a=2.5$ -- the same applies for $b$ -- and has two poles at which the sign changes. 
If deviations in the tail region of a distribution function are to be weighted more heavily, then stress parameters greater than zero are a suitable choice.
Focusing only on integer values for the stress parameter, 
the result for $a = 2$ is surprising. 
Since the risk function approaches infinity for these values, the associated weighting function and the corresponding statistic should not be used.
This result is in contrast to the excerpts from annonymous scripts found during our research that suggest these statistics.
Because of the symmetry, the stress parameters for $a = 1$ and $a = 4$ are equivalent with respect to the risk function. Only for $a\geq 5$ can marginal improvements be achieved. 
However, in the preliminary investigations, for large exponents and small samples of financial data, the evaluation of the corresponding statistics became numerically difficult.
Therefore, we suggest using statistics with $a = 1$ as the basis for a goodness of fit test, especially for the tail of a distribution or for determining the tail region of a parent distribution. \\

In finance, it is customary to represent losses by a coordinate transformation $z = -x$ in positive values and then perform the risk assessment at high quantiles in the upper tail. Regarding a goodness of fit test or a tail-detection method, this transformation is covered by lemma \ref{lemmasym}.
Therefore, in Section \ref{CV}, we further investigate the upper tail statistics $AU_n^2 = \hat R_{n,0,1}$ for stress parameter $b=1$.
\section{Model of the tail of a distribution}\label{GPD}
A theorem in extreme value theory, that goes back to \citet{gnedenko43}, \citet{balkema74} and \citet{pickands75},  
states that for a broad class of distributions, the distribution of the excesses over a threshold converges to a GPD, if the threshold is sufficiently large.\\

The GPD is usually expressed as a two-parameter distribution and has the following distribution function:
\begin{flalign}\label{GPDDF}
F(x) & = 1-\left(1+\xi\frac{x}{\sigma} \right)^{-\frac{1}{\xi}},
\end{flalign}
where $\sigma$ is a positive scale parameter and $\xi$ is a shape parameter. The density function is
\begin{flalign}\label{GPDPF}
f(x) & = \frac{1}{\sigma}\left(1+\xi\frac{x}{\sigma} \right)^{-\frac{1+\xi}{\xi}},
\end{flalign}
with support $0\leq x< \infty$ for $\xi\geq 0$ and $0\leq x\leq -\frac{\sigma}{\xi}$ when $\xi< 0$. The mean and variance are ${\text E}[x] = \frac{\sigma}{1-\xi}$ and ${\text{Var}}[x] = \frac{\sigma^2}{(1-\xi)^2(1-2\xi)} $, respectively; thus, the mean and variance of the GPD are positive and finite only for $\xi < 1$ and $\xi < 0.5$, respectively. For special values of $\xi$, the GPD leads to various other distributions. When $\xi = 0, -1$, the GPD becomes an exponential or a uniform distribution, respectively. For $\xi>0$, the GPD has a long tail to the right and is a reparameterized version of the usual Pareto distribution. Several areas of applied statistics have used the latter range of $\xi$ to model data sets that exhibit this form of a long tail. \\

Since the GPD was introduced by \citet{pickands75}, numerous theoretical advancements and applications have followed \citep{davison84, smith84, smith85,vanmontfort85,hosking87,davison90,choulakian01}. Its applications include use in the analysis of extreme events in hydrology, as a failure-time distribution in reliability studies and in the modeling of large insurance claims. Numerous examples of applications can be found in \citet{embrechts03} and the studies listed therein. The GPD is also increasingly used in the financial and banking sectors. Especially in the assessment of risks based on high quantiles, the GPD is one of the proposed distributions for modeling the tail of an unknown parent distribution \citep{basel09}.\\

The preferred method in the literature for estimating the parameters of the GPD is the well-studied maximum likelihood method \citep{davison84,smith84,smith85,hosking87}. \citet{choulakian01} stated that it is theoretically possible to have data sets for which no solution to the likelihood equations exists, and they concluded that, in practice, this is extremely rare. In many practical applications, the estimated shape parameter $\hat\xi$ is in the range between -0.5 and 0.5, and a solution to the likelihood equations exists \citep{hosking87,choulakian01}. For practical and theoretical reasons, these authors limit their attention to this range of values. We adapted the GPD as a model for the tail of different parent distributions applicable to finance and banking (Section \ref{TD}, Fig.\ \ref{xiparameter}) and also found that the value of $\hat\xi$ falls within this range. Therefore, we also focus mainly on the range $-0.5<\xi<0.5$ and use the standardized maximum likelihood method to estimate the parameters of the GPD based on the data. Furthermore, we check the behavior of the critical values in the next section only for $\xi = 0.9$. This is a test for the rare case in which $\xi$ approaches 1.0 and thus the expected value approaches infinity.
\section{Critical values of the test statistics}\label{CV}
In goodness of fit tests, the critical value is the cut-off value used to decide whether the null hypothesis $H_0$ -- the sample $x_1, x_2, \ldots , x_n$ originating from a specific distribution $F(x)$ -- is rejected at a given significance level or not. The critical values are generally dependent on the underlying distribution function if the parameters of the distribution need to be estimated from the data. Furthermore, they are usually still dependent on the data size $n$. Recall that the case in which the parameters of the assumed distribution are unknown is referred to as case 3, while the case in which the parameters are completely specified is referred to as case 0, according to \citet{stephens71, stephens76}. In the following, Eq.\ (\ref{GPDDF}) is assumed to represent the tail model sought. According to \citet{choulakian01}, two further cases should be distinguished in this context: case 1, in which the shape parameter $\xi$ is known and the scale parameter $\sigma$ is unknown, and case 2, in which the shape parameter $\xi$ is unknown and the scale parameter $\sigma$ is known. Case 3, in which both parameters are unknown, describes the important and most likely situation to arise in practice.
In this case, the critical values also depend on the value of the estimated shape parameter $\hat \xi$ of the GPD. \\

To obtain critical values for carrying out the goodness of fit test, various methods are available. \citet{choulakian01} list tables of asymptotic percentage points for the Anderson-Darling statistic $A_n^2$ and the Cram\'er-von Mises statistic $W_n^2$. They calculated the percentage points by using the asymptotic theory of statistic tests and by following a procedure described earlier by, for instance, \citet{stephens76}. \citet{heo13} performed Monte Carlo simulations to obtain the critical values for the Anderson-Darling statistic $A_n^2$ and the modified Anderson-Darling statistic $AU_n^2$. They provided regression equations that depend on the sample size $n$, shape parameter $\xi$ and the significance level $p$ to calculate approximations of the critical values of the statistics for the GPD \citep[Table 4 therein]{heo13}. For the Anderson-Darling statistic $A_n^2$, the critical values obtained by both methods can be compared. Notably, slight discrepancies exist between the results of the authors. \\

In the goodness of fit tests, the significance level $p$ is usually set beforehand in the range $ 0.001 <p <0.5 $ so that the available tables of the critical values usually cover only this range. Since we are interested in the minimum of the test statistic $AU_n^2 = \hat R_{n,0,1}$ in the context of tail detection, the critical values for significance levels in the range $ 0.5 <p <1.0 $ are required.\\

With the minimum of the test statistic $AU_n^2$, the threshold $u$ -- at which the tail starts -- and the tail model are found. The tail model above the threshold $u$ consists of the GPD with the estimated parameters $\hat\sigma$ and $ \hat\xi$.
To evaluate the quality of the adaptation of the GPD as a tail model to the data that lie above the threshold $u$, the standard goodness of fit tests according to Cram\'er-von Mises and Anderson-Darling are performed.
The test quantities $W_n^2$ and $A_n^2$ are expected to also be very small.
Consequently, critical values corresponding to significance levels of $0.5<p<1.0$ are needed. 
These critical values are currently not available and must also be predetermined.\\

Comprehensive Monte Carlo simulations are performed to calculate the required critical values for the GPD. 
For this purpose, data sets $x_i$ of finite length $n$ up to 2000 are generated using the GPD with given shape parameter $\xi$. The shape parameters assigned in this simulation experiment are shown in Table \ref{cvtab}. The scale parameter $\sigma$ is set to 1.\\

For a given shape parameter, 5m sets of data with sample size $n$ are generated. The distribution parameters are then estimated from each generated random sample, and the test statistics $W_n^2$, $A_n^2$ and $AU_n^2$ of each data set are calculated. These 5m statistic values are subsequently ranked in ascending order, and the critical values for the desired significance levels listed in Table \ref{cvtab} are determined for each combination of $n$ and $\xi$. During our analysis, the critical values were observed to converge quickly to an asymptotic limit, a peculiarity of the statistics $W_n^2$, $A_n^2$ and $AU_n^2$ that can be found in many applications \citep{choulakian01}.\\

In the following, we use the asymptotic limits $W^2$, $A^2$ and $AU^2$ shown in Table \ref{cvtab} as approximations of the critical values for finite $n$.
The Monte Carlo simulations show that the critical values listed in Table \ref{cvtab} can be used with good accuracy for data sets with small samples sizes. However, as suggested in \citet {choulakian01}, the samples should be at least $n> 25$ in size. Later, we will see that our results in Section \ref{Simulation} indicate the same lower limit.\\

For some combinations of significance levels $p$ and shape parameter $\xi$, the critical values for $W_n^2$ and $A_n^2$ can be compared with the asymptotic percentage points calculated from the asymptotic theory shown in Table 2 of \citet{choulakian01}. Generally, the relative deviation of the results is less than 1\%. Only for small values of $\xi$ and very small significance levels $p$ does the relative deviation increase on the order of 10\%.
\begin{sidewaystable}
	\captionsetup{labelsep=newline, justification=RaggedRight, singlelinecheck=false, labelfont = bf}
	\caption{Approximation of critical values for the statistics $W_n^2$, $A_n^2$ and $AU_n^2$ for the GPD when both parameters have to be estimated.}\label{cvtab}
	\renewcommand{\arraystretch}{1.1}
	\begin{tabular}{rc|rrrrrr|rrrrrrr}
		\hline
     &                             & \multicolumn{6}{|l}{\textbf{Tail-Detection}} & \multicolumn{7}{|l}{\textbf{Goodness-Of-Fit Tests}}   \\[6pt]
$\xi$& Statistic \textbackslash $\;p$  & 0.950 & 0.900& 0.850& 0.800&	0.750&	0.500 &	0.250&	0.100&	0.050&	0.025&	0.010&	0.005&	0.001 \\ \hline
-0.5 & $W^2$                       & 0.027 & 0.032 & 0.037 & 0.041 & 0.045 & 0.068 & 0.104 & 0.155 & 0.194 & 0.236 & 0.293 & 0.336 & 0.439 \\
	 & $A^2$                       & 0.203 & 0.239 & 0.269 & 0.296 & 0.321 & 0.459 & 0.674 & 0.965 & 1.195 & 1.435 & 1.765 & 2.018 & 2.621 \\       	
	 & $AU^2$                      & 0.085 & 0.100 & 0.112 & 0.123 & 0.134 & 0.191 & 0.277 & 0.389 & 0.476 & 0.565 & 0.686 & 0.778 & 0.995 \\[6pt]     
-0.4 & $W^2$                       & 0.026 & 0.031 & 0.036 & 0.040 & 0.044 & 0.065 & 0.100 & 0.147 & 0.185 & 0.223 & 0.276 & 0.317 & 0.414 \\
	 & $A^2$                       & 0.198 & 0.234 & 0.262 & 0.288 & 0.313 & 0.445 & 0.650 & 0.926 & 1.146 & 1.373 & 1.686 & 1.927 & 2.502 \\       	
	 & $AU^2$                      & 0.082 & 0.097 & 0.109 & 0.119 & 0.130 & 0.184 & 0.265 & 0.371 & 0.453 & 0.536 & 0.650 & 0.737 & 0.945 \\[6pt] 
-0.3 & $W^2$                       & 0.025 & 0.030 & 0.035 & 0.038 & 0.042 & 0.063 & 0.095 & 0.140 & 0.175 & 0.212 & 0.261 & 0.300 & 0.392 \\
	 & $A^2$                       & 0.194 & 0.228 & 0.255 & 0.280 & 0.304 & 0.431 & 0.627 & 0.890 & 1.099 & 1.315 & 1.610 & 1.839 & 2.388 \\       	
	 & $AU^2$                      & 0.080 & 0.094 & 0.106 & 0.116 & 0.126 & 0.177 & 0.254 & 0.355 & 0.432 & 0.511 & 0.618 & 0.701 & 0.897 \\[6pt] 	 
-0.2 & $W^2$                       & 0.025 & 0.030 & 0.034 & 0.037 & 0.041 & 0.060 & 0.091 & 0.133 & 0.166 & 0.200 & 0.246 & 0.282 & 0.368 \\
	 & $A^2$                       & 0.190 & 0.223 & 0.249 & 0.273 & 0.297 & 0.418 & 0.606 & 0.855 & 1.052 & 1.256 & 1.537 & 1.752 & 2.275 \\       	
	 & $AU^2$                      & 0.078 & 0.092 & 0.103 & 0.113 & 0.122 & 0.171 & 0.245 & 0.340 & 0.413 & 0.487 & 0.588 & 0.666 & 0.851 \\[6pt] 
-0.1 & $W^2$                       & 0.024 & 0.029 & 0.033 & 0.036 & 0.040 & 0.058 & 0.087 & 0.127 & 0.157 & 0.189 & 0.233 & 0.266 & 0.348 \\
	 & $A^2$                       & 0.186 & 0.218 & 0.244 & 0.267 & 0.289 & 0.406 & 0.584 & 0.822 & 1.010 & 1.204 & 1.468 & 1.671 & 2.164 \\       	
	 & $AU^2$                      & 0.077 & 0.090 & 0.100 & 0.110 & 0.119 & 0.166 & 0.236 & 0.326 & 0.396 & 0.467 & 0.563 & 0.636 & 0.811 \\[6pt] 	  
 0.0 & $W^2$                       & 0.024 & 0.028 & 0.032 & 0.035 & 0.039 & 0.056 & 0.084 & 0.121 & 0.150 & 0.180 & 0.221 & 0.253 & 0.327 \\
	 & $A^2$                       & 0.183 & 0.214 & 0.238 & 0.261 & 0.282 & 0.395 & 0.565 & 0.791 & 0.970 & 1.153 & 1.406 & 1.602 & 2.062 \\       	
	 & $AU^2$                      & 0.075 & 0.088 & 0.098 & 0.107 & 0.116 & 0.161 & 0.229 & 0.315 & 0.381 & 0.449 & 0.540 & 0.611 & 0.777 \\[6pt] 	 
 0.1 & $W^2$                       & 0.023 & 0.027 & 0.031 & 0.034 & 0.037 & 0.054 & 0.081 & 0.116 & 0.143 & 0.171 & 0.209 & 0.239 & 0.309 \\
	 & $A^2$                       & 0.180 & 0.210 & 0.234 & 0.256 & 0.276 & 0.385 & 0.549 & 0.765 & 0.935 & 1.109 & 1.348 & 1.533 & 1.975 \\       	
	 & $AU^2$                      & 0.074 & 0.087 & 0.097 & 0.105 & 0.114 & 0.158 & 0.223 & 0.306 & 0.369 & 0.434 & 0.521 & 0.588 & 0.746 \\[6pt] 
 0.2 & $W^2$                       & 0.023 & 0.027 & 0.030 & 0.034 & 0.037 & 0.053 & 0.078 & 0.111 & 0.137 & 0.164 & 0.200 & 0.228 & 0.294 \\
	 & $A^2$                       & 0.177 & 0.206 & 0.230 & 0.251 & 0.271 & 0.376 & 0.534 & 0.741 & 0.903 & 1.070 & 1.298 & 1.474 & 1.889 \\       	
	 & $AU^2$                      & 0.073 & 0.085 & 0.095 & 0.104 & 0.112 & 0.155 & 0.218 & 0.298 & 0.359 & 0.421 & 0.505 & 0.569 & 0.720 \\[6pt] 	 
 0.5 & $W^2$                       & 0.022 & 0.026 & 0.029 & 0.032 & 0.034 & 0.049 & 0.072 & 0.101 & 0.124 & 0.148 & 0.179 & 0.204 & 0.263 \\
	 & $A^2$                       & 0.171 & 0.199 & 0.220 & 0.240 & 0.259 & 0.356 & 0.499 & 0.686 & 0.831 & 0.980 & 1.183 & 1.339 & 1.715 \\       	
	 & $AU^2$                      & 0.071 & 0.083 & 0.092 & 0.101 & 0.108 & 0.149 & 0.208 & 0.283 & 0.340 & 0.398 & 0.477 & 0.536 & 0.678 \\[6pt] 
 0.9 & $W^2$                       & 0.021 & 0.024 & 0.027 & 0.030 & 0.033 & 0.046 & 0.067 & 0.094 & 0.115 & 0.136 & 0.165 & 0.187 & 0.240 \\
	 & $A^2$                       & 0.166 & 0.192 & 0.213 & 0.232 & 0.249 & 0.339 & 0.472 & 0.641 & 0.772 & 0.905 & 1.087 & 1.229 & 1.568 \\       	
	 & $AU^2$                      & 0.071 & 0.082 & 0.091 & 0.099 & 0.107 & 0.146 & 0.204 & 0.277 & 0.333 & 0.389 & 0.465 & 0.523 & 0.661 \\ \hline	    		
	\end{tabular}
\end{sidewaystable}
\newpage
\section{Detecting the begin of the tail} \label{TD}
In this section, the procedure for detecting the beginning of the upper tail of an unknown parent distribution based on the statistic $AU_n^2 \; (= \hat R_{n,0,1})$ is described, cf.\ Eq. (\ref{LTSEEQb1}). Due to lemma \ref{lemmasym}, the procedure described below can also be used to detect the lower tail of an unknown parent distribution if we change the sign of the measured data.\\

Given a data set, the detection of the tail essentially involves determining the optimal threshold value $u$ at which the tail of the unknown parent distribution starts and the GPD can be used as a model.
In the descending-order time series, the optimal threshold $u$ is a special data point $x_{(k)}$, with $k\leq n$, and the ordered data points $x_{(1)}, x_{(2)}, \ldots , x_{(k)}$ are values that come from the tail of the parent distribution. The latter data set is used to estimate the unknown parameters $\xi$ and $\sigma$ of the GPD (see Eq.\ (\ref{GPDDF})). 
\subsection{Procedure}\label{PROCEDURE}
In the general case, the tail model is determined as follows:
\begin{enumerate}
	\item Sort the random sample taken from an unknown parent distribution in descending order: $x_{(1)} \geq x_{(2)} \geq \ldots \geq x_{(n)}$.
	\item Let $k =2,\ldots, n$, and find for each $k$ the estimates $\hat{\xi}_k$ and $\hat{\sigma}_k$ of the parameters of the GPD $F(x)$ (see Eq.\ (\ref{GPDDF})), as described in Section \ref{GPD}. Note: For numerical reasons, we start at k = 2.
	\item Calculate the probabilities $F(x_{(i)})$ for $i=1,\ldots,k$, and determine the statistics $AU_k^2$, $W_k^2$ and $A_k^2$, cf.\ Eq.\ (\ref{LTSEEQb1}), (\ref{CVM}) and (\ref{ADTS}), respectively. 
	\item Find the index $k^*$ of the minimum of the statistics $AU_k^2$.
\end{enumerate}
Then, the optimal threshold value is estimated by $\hat u = x_{(k^*)}$, and the model of the tail of the unknown parent distribution is given by the GPD with parameter estimates $\hat{\xi}_{k^*}$ and $\hat{\sigma}_{k^*}$. \\

Table \ref{cvtab} gives the critical values for the statistics under consideration to perform the goodness of fit test for a given level of significance $p$. In case 3, where $\xi$ must be estimated, the table should be entered at $\hat{\xi}_{k^*}$. If $\hat{\xi}_{k^*} < -0.5$, the table should be entered at $\xi = -0.5$. Critical values for other values of $\xi$ can be obtained by interpolation \citep{choulakian01}. The table can also be used to deduce the $p$-value from the estimated parameter $\hat{\xi}_{k^*}$ and the statistics. The $p$-value is then a measure of the fallacy if the estimated GPD $F(x;\; \hat{\xi}_{k^*}, \hat{\sigma}_{k^*} )$ is rejected as a tail model. This leads to the idea that the $p$-value in a decision should be as large as possible for the GPD not to be rejected. We will return to this point later in the applications.
\subsection{Ideal cases}\label{IDCA}
For the selection of parent distributions $H(x)$ listed in Table \ref{parentdist}, we first describe the procedure. The selected distributions correspond to the distributions commonly used in the finance and insurance industries.
To study the procedure and some features of the method, ideal situations are prepared.
\begin{table}
	\captionsetup{labelsep=newline, justification=RaggedRight, singlelinecheck=false, labelfont = bf}
	\caption{Parent Distributions.}\label{parentdist}
	\renewcommand{\arraystretch}{1.1}
	\begin{tabular}{lccc}
		\hline
		Parent Distribution & \multicolumn{3}{l}{Parameter}     \\[0pt]
		             & Location & Scale      & Shape     \\[0pt]
		     $H(x)$  & {$\mu $} & {$\sigma$} & {$\xi$ }  \\ \hline
		Lognormal    & 0.0      & 1.0        &  -        \\
		Normal       & 0.0      & 1.0        &  -        \\
		Generalized Extreme Value (GEV) & 0.0 & 1.0 & 0.5 \\
		Generalized Pareto Distribution (GPD)       & 0.0 & 1.0 & 0.5 \\
		Exponential  & -        & 1.0        &   -       \\ \hline
\end{tabular}
\end{table}
The ideal case for each parent distribution is prepared by setting $\frac{k}{n} = H(x_k)$ for $k =1, \ldots, n$ and determining the quantiles $x_k$. This data set is then used as the input for the procedure described in Section \ref{PROCEDURE}. Data records of length $n = $ 50, 100, 500, 1k, 2k, 3k, 5k, 10k, 50k and 100k are examined. \\

\begin{figure}[htbp]
	\centering
	\captionsetup{labelfont = bf}
	\includegraphics[width=0.975\textwidth]{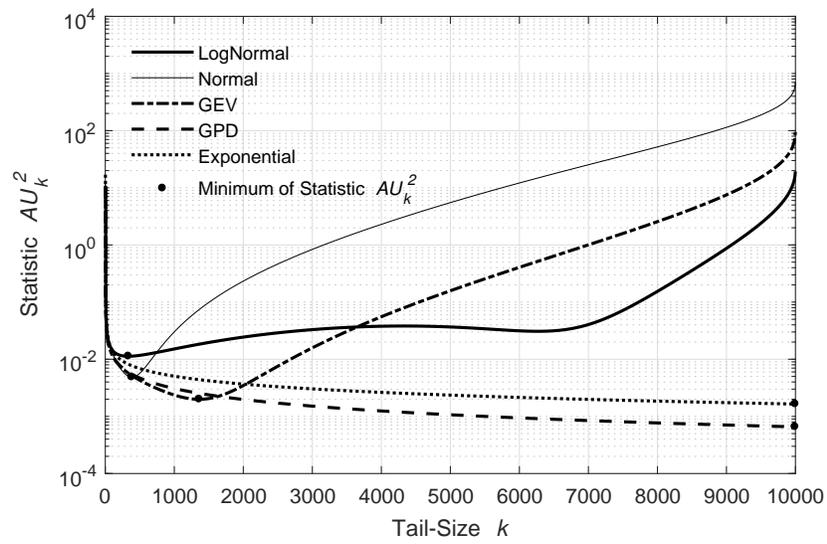}
	\begin{quote}
		\caption[Sample-Size 100]{\label{idealcase} How the statistic $AU_k^2$ behaves as a function of the growing tail length $k$ for the parent distributions from Table \ref{parentdist}. Bullet points indicate the respective minimums of statistics. These minimum values show the optimal data set length that should be used to determine the parameters of the GPD.}
	\end{quote}
\end{figure}

Fig.\ \ref{idealcase} shows, for $n =$ 10k, the value of the statistic $AU_k^2$ as the length $k$ of the tail increases until the total data length comprises.
Except in the case of the GPD and the exponential distribution, the statistics of the remaining distributions will show a minimum at less than 20\% of the total data length $n$. If the GPD is used as the parent distribution, the whole data set must be used to estimate the parameters of the GPD as the tail model as accurately as possible. Fig.\ \ref{idealcase} shows that in the case of the GPD, the statistic approaches zero as the length of the tail reaches its maximum. Because the exponential distribution is a special GPD with $\xi = 0$, the same applies to this distribution. \\

\begin{figure}[htbp]
	\centering
	\captionsetup{labelfont = bf}
	\includegraphics[width=0.975\textwidth]{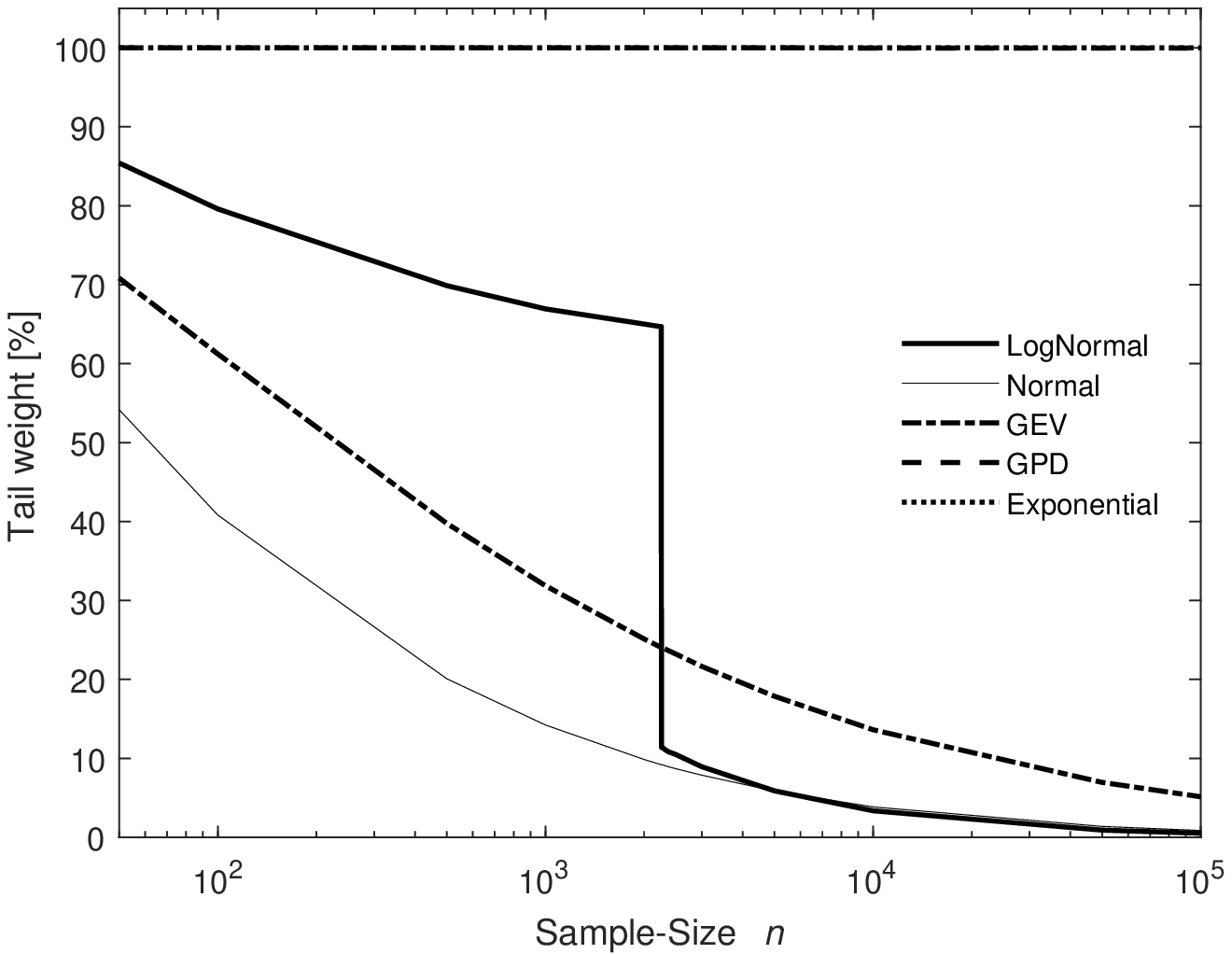}
	\begin{quote}
		\caption[Sample-Size 100]{\label{tailweight} Dependence of the proportion of tail points on the total amount of data. The GPD and the exponential distribution always use the whole dataset (tail weight 100\%), so the two results are indistinguishable here. }
	\end{quote}
\end{figure}

More detailed insight regarding the remaining distributions is given by Fig.\ \ref{tailweight}.
As the total number of data points increases, the proportion of data points used to model the tail decreases. Conversely, the lower the total number of data points, the larger the proportion needed to model the tail. In addition, as with the lognormal distribution, a local minimum may still exist near the global one, both of which could change their characteristics as the total amount of data increases.
This is due to the fact that in this case, before and after the mode of the parent distribution, local minima of the statistics are calculated. One of these minima is the global minimum for the dataset. With small amounts of data, the statistics show that it is more advantageous to also use data beyond the mode to adapt the model. For large data sets, the advantage disappears and only data belonging to the tail area of the parent distribution -- before the mode -- are used.
This result may indicate that, generally, no simple relationship exists between the optimal length of the tail and the total amount of data. Previous reports have already carried out studies on the favorable choice of threshold in the finance and insurance fields \citep{mcneil97, moscadelli04, dutta07}, revealing that the preferred tail length for the data series analyzed in those fields comprises approximately 10\% to 15\% of the total amount of data available. Based on our results shown in Fig.\ \ref{tailweight}, we conclude that this proportion should not be regarded as a rule of thumb in future works without a case-by-case examination.\\

\begin{figure}[htbp]
 	\centering
 	\captionsetup{labelfont = bf}
 	\includegraphics[width=0.975\textwidth]{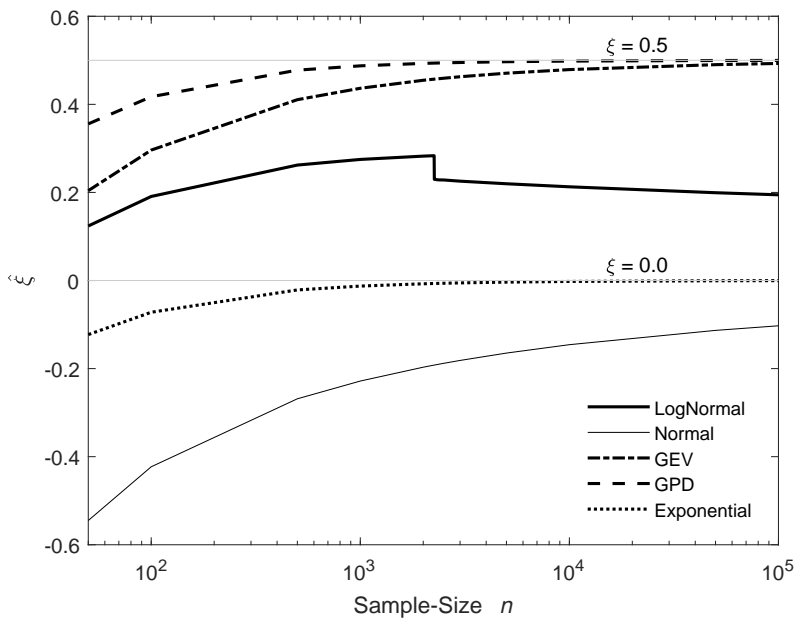}
 	\begin{quote}
 		\caption[Sample-Size 100]{\label{xiparameter} Dependence of the estimated parameter $\xi$ on the total amount of data.}
 	\end{quote}
\end{figure}

Fig.\ \ref{xiparameter} shows the dependency of the parameter $\xi$, estimated at the minimum of the statistics $AU_k^2$, on the total length $n$ of the data series. While in the case of the GEV, the GPD and the exponential distribution, the parameter $\xi$ well converges to the true value (cf.\ Table \ref{parentdist}), in the case of the lognormal and the normal distribution, no such limit value is yet apparent.

\subsection{Monte Carlo simulations}\label{Simulation}
\begin{figure}[htbp]
	\centering
	\captionsetup{labelfont = bf}
	\includegraphics[width=0.975\textwidth]{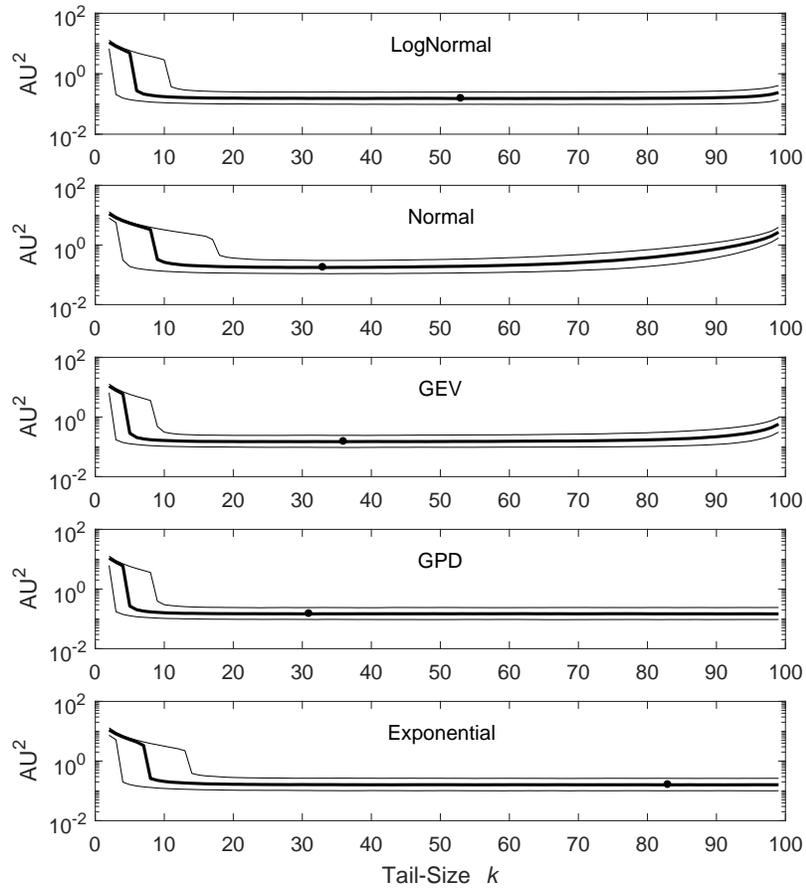}
	\begin{quote}
		\caption[Sample-Size 100]{\label{sample100} For the various parent distributions, the mean of the statistic $AU^2_k$ is displayed (bold lines) when the tail size $k$ used reaches the sample size $n =100$. The $1 \sigma$ confidence interval (thin lines) encloses this mean. The minimum is indicated by a bullet point.}
	\end{quote}
\end{figure}

We now examine how the procedure works when considering the average of many sets of data. Monte Carlo simulations were performed for each distribution listed in Table \ref{parentdist} such that a sample $x_1, x_2, \ldots, x_n$ of size $n$ was generated for an assumed parent distribution with given parameters. As in the previous section, we analyzed different sample lengths for $n =$ 50, 100, 500 and 1k. For the given parameter set, 10k sets of data were generated for each sample size. The proposed procedure for detecting the beginning of the upper tail was then applied for each data set. For each $k =2,\ldots, n$, an empirical distribution of the statistic $AU^2_k$ exists, which we evaluated with regard to the mean value $\overline{AU}^{\;2}_k$ and a $1 \sigma$ confidence interval. The confidence interval thus narrows the range, with approximately 67\% of the realizations of the statistic $AU_k^2$. Furthermore, we determined the minimum of $\overline{AU}^{\;2}_k$ with respect to the size of the tail $k$. Fig.\ \ref{sample100} shows the results for the data set with $n = 100$ as an example. As in the ideal situation (see Section \ref{IDCA}), we observe the same graphical progress.\\

In Section \ref{CV}, we noticed that the statistic $AU^2_k$ for tail size $k \leq 25$ converges only poorly and that the distribution of $AU^2_k$ is correspondingly wide, cf.\ the confidence interval shown in Fig.\ \ref{sample100}. For $k>25$, the mean value of $AU^2_k$ approaches a minimum. In the case of the GPD and exponential distribution, this must occur at the edge, but due to the finite Monte Carlo simulation, the smallest fluctuations remain. Thus, the minimum has already been detected before, with $k^*\leq n$. For the lognormal, normal and GEV distribution, the mean of the statistic rises significantly after the minimum at $k^*$ has been passed and $k \rightarrow n$. In these one-humped distributions, the modal value is designated by a certain large $k_m \geq k^*$. From this $k_m$, the mean value of the statistic $AU^2_k$ very clearly increases as $k$ approaches the sample size $n$.
\section{Examples}\label{examples}
In this section, we consider a specific example with a known parent distribution in detail and then discuss the results when our procedure is applied to the MSCI World Index.
The first example shows the application of the procedure to a distribution function that is commonly used in finance when the modeling of stock returns is required. For an exemplary single time series, the tail model is determined. The example also shows how the tail model can be evaluated using standard goodness of fit tests.
The second example then shows the application of the method to real data, i.e.,\ to time series of which the true distribution is generally unknown.
\subsection{Single data row with known parent distributions}\label{SimulationSingleData}
In the first example, the data are the $n = 200$ values generated from a lognormal distribution with parameters specified in Table \ref{parentdist}. Our goal is to first find within that data the subset of data that comes from the tail of the parent distribution. The parameters of the GPD -- as a model for the tail of the parent distribution -- are then estimated based on the data subset. \\

\begin{figure}[htbp]
	\centering
	\captionsetup{labelfont = bf}
	\includegraphics[width=0.975\textwidth]{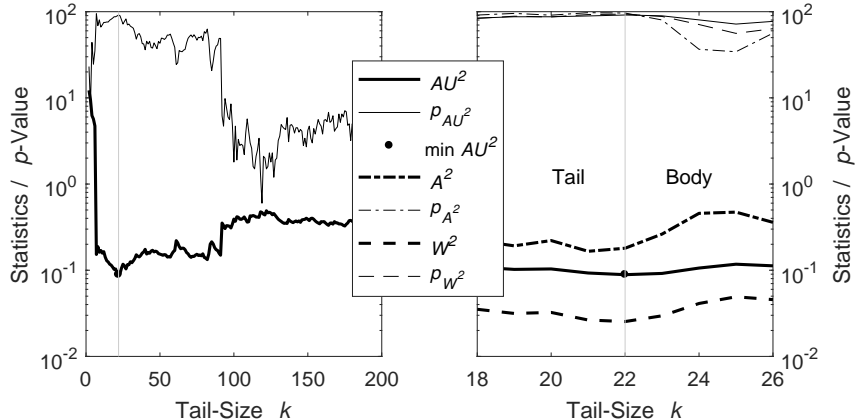}
	\begin{quote}
		\caption[Sample-Size 100]{\label{example1p1} Application of the procedure for detecting the beginning of the tail for $n=200$ data points generated with the lognormal distribution. Left figure: For the entire sample $k=1,\ldots,n$, the statistic $AU_k^2$ (bold line) and the $p_{AU^2}$-value (thin line; in percent) are shown. The minimum of $AU_k^2$ is indicated by a bullet point and a thin vertical line. Right figure: The relevant area around the minimum of $AU_k^2$ is shown in an enlargement. Additionally marked are the Anderson-Darling ($A_k^2$) and the Cram\'er-von Mises ($W_k^2$) statistic, as well as the corresponding $p$-values greater than 90\%.}
	\end{quote}
\end{figure}

The procedure proposed in Section \ref{PROCEDURE} results in a minimum value of the upper tail statistic of $AU_{22}^2 = 0.0888$, indicating that in this case, 22 data points are taken from the tail of the parent distribution. Fig.\ \ref{example1p1} (left) shows the value of the statistic $AU_k^2$ as the tail length $k$ increases. In addition, the $p$-value of the statistics -- determined via a Monte Carlo simulation -- is shown for each $k$. On the right side of Fig.\ \ref{example1p1}, an enlargement of the relevant range around the minimum of the upper tail statistic at $k^* = 22$ is shown. Furthermore, the values of the other two statistics $W_k^2$ and $A_k^2$, as well as the corresponding $p$-values, are plotted versus increasing $k$.\\

Table \ref{example1results} summarizes the results of modeling the tail of the lognormal distribution.
In addition to the parameters of the GPD, as the model used for the tail, the results of the three statistics used here are listed. For the upper tail statistic $AU_k^2$, the determined minimum value is given, cf.\ Fig.\ \ref{example1p1} (right). The two other statistics $W_k^2$ and $A_k^2$ are used to evaluate the goodness of the fit. Based on the value of the respective statistics and the estimated parameter $\hat\xi$, the corresponding $p$-value can be determined from Table \ref{cvtab} -- for all statistics, we found $p$-values greater than 0.90. Via an additional Monte Carlo simulation, we can provide more accurate estimates of the $p$-values listed in Table \ref{example1results} for this example. All three statistics indicate that the tail model was adapted with sufficiently high quality.\\

\begin{figure}[htbp]
	\centering
	\captionsetup{labelfont = bf}
	\includegraphics[width=0.975\textwidth]{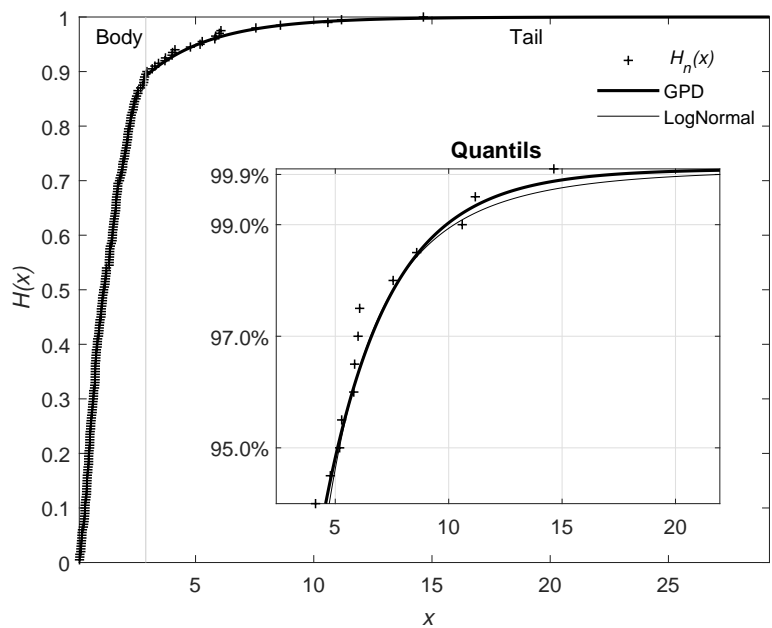}
	\begin{quote}
		\caption[Sample-Size 100]{\label{example1p2} Separation of the body and the tail of the empirical distribution $H_n(x)$ at a threshold $u= x_{(22)} = 2.85$. The inner picture shows the tail region and the fitted GPD (bold line) in comparison to the empirical and the true lognormal distribution (thin line). Even at this magnification, the two are difficult to distinguish. As an example, some common quantiles are drawn.  }
	\end{quote}
\end{figure}

\begin{table}
	\captionsetup{labelsep=newline, justification=RaggedRight, singlelinecheck=false, labelfont = bf}
	\caption{Goodness of fit test for the tail model.}\label{example1results}
	\renewcommand{\arraystretch}{1.1}
	\begin{tabular}{lcc}
		\hline
		Parent Distribution  	& \multicolumn{2}{l}{Lognormal}   	\\[6pt]
		Tail Model   			& \multicolumn{2}{l}{GPD}   	 	\\[0pt]
		Parameter    			& $\hat{\xi}$    &  0.037   		\\[0pt]
		& $\hat{\sigma}$ &  2.824   		\\[6pt]
		Test		 			& Statistic   	 & {$p$-Value}      \\ \hline
		$W_{22}^2$        			& 0.0254         & 0.937            \\
		$A_{22}^2$        			& 0.1800         & 0.974            \\
		$AU_{22}^2$       			& 0.0888         & 0.915            \\ \hline
	\end{tabular}
\end{table}
Fig.\ \ref{example1p2} shows the empirical distribution of the data and illustrates the quality of the fit. From the threshold $u= x_{(22)} = 2.85$, indicated by the 22 data points of the sorted data set, the set is divided into two subsets. The data below the threshold come from the body, while that above the threshold, from the tail of the parent distribution $H(x)$. The subset of data belonging to the tail of the parent distribution is used to estimate the parameters of the GPD via the maximum likelihood method. The inserted graphic shows an enlargement of the tail region for those quantiles typically used in practice.
In this constructed example, it is possible to compare the graph of the GPD and the lognormal distribution at high quantiles. The deviations between the graphs are comparatively small in this case.
\subsection{MSCI World Index}
The MSCI World Index is a market capital weighted stock market index covering stocks of all the developed markets in the world, as defined by MSCI Inc.\, formerly Morgan Stanley Capital International. Below, for the period from 31.12.1969 to 31.12.2017, we consider the weekly, monthly and annual closing price time series of the MSCI World Index (Bloomberg ticker code: MXWO) in USD. 
A number of products are available on the capital market, which are related to the MSCI World Index.
For an investor, in addition to the expected return, the risk associated with a product is usually interesting. An investor has various procedures and risk indicators available for assessing risk.
In the present study, we want to estimate the investment risk associated with the products by analysing the statistical behavior of the rate of change of the underlying index and the risk indicators derived from it.
We are particularly interested in the VaR and CVaR of the index; for definitions, cf.\ \citet[ch.\ 22]{hull17}. 
The model of the stochastic process usually assumed for the price $S$ of a non-dividend-paying stock is known as geometric Brownian motion. 
To determine the model parameters of the corresponding stochastic differential equation and to derive further risk indicators, the logarithmic rate of return of the stock price are considered. This model implies that the logarithmic rate of return is normally distributed, cf.\ \citet[ch.\ 14]{hull17}. Consistent with this assumption, we examine the logarithmic rate of return of the index and apply our method to determine the model for the tail. The model is then used to calculate the desired risk parameters VaR and CVaR for the confidence levels usually assumed in a risk report: 95.0\%, 97.0\%, 99.0\% and 99.9\%, cf.,\ e.g.,\ \citet{basel04}.\\

Generally, the risk indicators should be positive; thus, we change the sign of the logarithmic return rate. This corresponds to the coordinate transformation assumed in lemma \ref{lemmasym}. Hence, it is now possible to use the upper tail statistic $AU_k^2$ to determine the threshold $u=x_{(k^*)}$ for the transformed and sorted time series $x_{(k)}$ for $k=1,\ldots,n$ of the logarithmic return rates, which marks the beginning of the tail. For the three time series considered, we have $n = 2503$ (weekly sampling), $n = 575$ (monthly sampling) and $n = 47$ (annual sampling). 
\subsubsection{Weekly sampling}\label{msciweekly}
\begin{figure}[htbp]
	\centering
	\captionsetup{labelfont = bf}
	\includegraphics[width=0.975\textwidth]{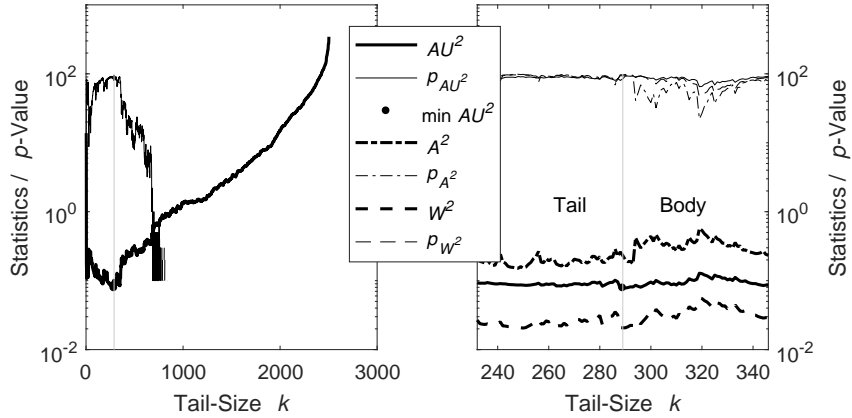}
	\begin{quote}
		\caption[Sample-Size 100]{\label{example2p1} Application of the procedure for detecting the beginning of the tail for $n=2503$ weekly logarithmic rates of return. Left figure: For the entire sample $k=1,\ldots,n$, the statistic $AU_k^2$ (bold line) and the $p_{AU^2}$-value (thin line; in percent) are shown. The minimum of $AU_k^2$ is indicated by a bullet point and a thin vertical line. Right figure: The relevant area around the minimum of $AU_k^2$ is shown in an enlargement. Additionally marked are the Anderson-Darling ($A_k^2$) and the Cram\'er-von Mises ($W_k^2$) statistic, as well as the corresponding $p$-values greater than 90\%.}
	\end{quote}
\end{figure}

First, we take a closer look at the logarithmic rates of returns of the weekly sampled data. Fig.\ \ref{example2p1} shows the results after applying the procedure to detect the beginning of the tail for the sorted data. For the statistic $AU^2_k$, the minimum occurs at $k^* = 289$. Thus, the threshold from which the tail model can be adapted to the exceeding data is given by the value $u =  x_{(289)} = 0.020$. 
With the data points that exceed this threshold, the parameters of the GPD are determined using the maximum likelihood method. At the same time, the statistics $W_k^2$ and $A_k^2$ and their $p$-values are calculated to perform the goodness of fit test. The results of all statistics and $p$-values, as well as the parameters of the adapted GPD, are summarized in Table \ref{example2results}. All statistics have very low values. Consequently, consistent with Table \ref{cvtab}, the $p$-value is high, which makes rejecting the assumption -- the GPD is the best model for the tail of the unknown parent distribution -- seemingly the wrong decision. \\

\begin{figure}[htbp]
	\centering
	\captionsetup{labelfont = bf}
	\includegraphics[width=0.975\textwidth]{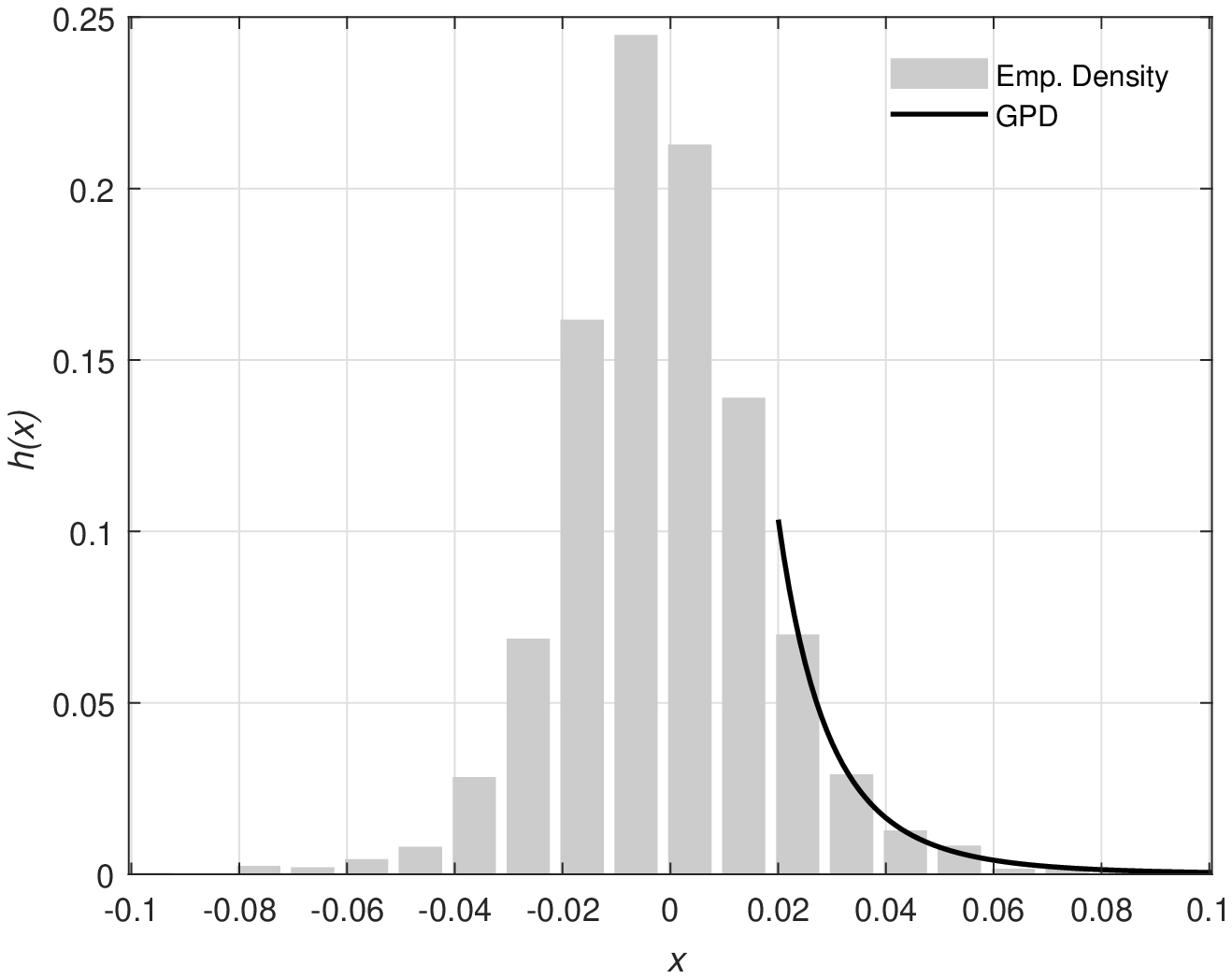}\\[-6pt]
	\includegraphics[width=0.975\textwidth]{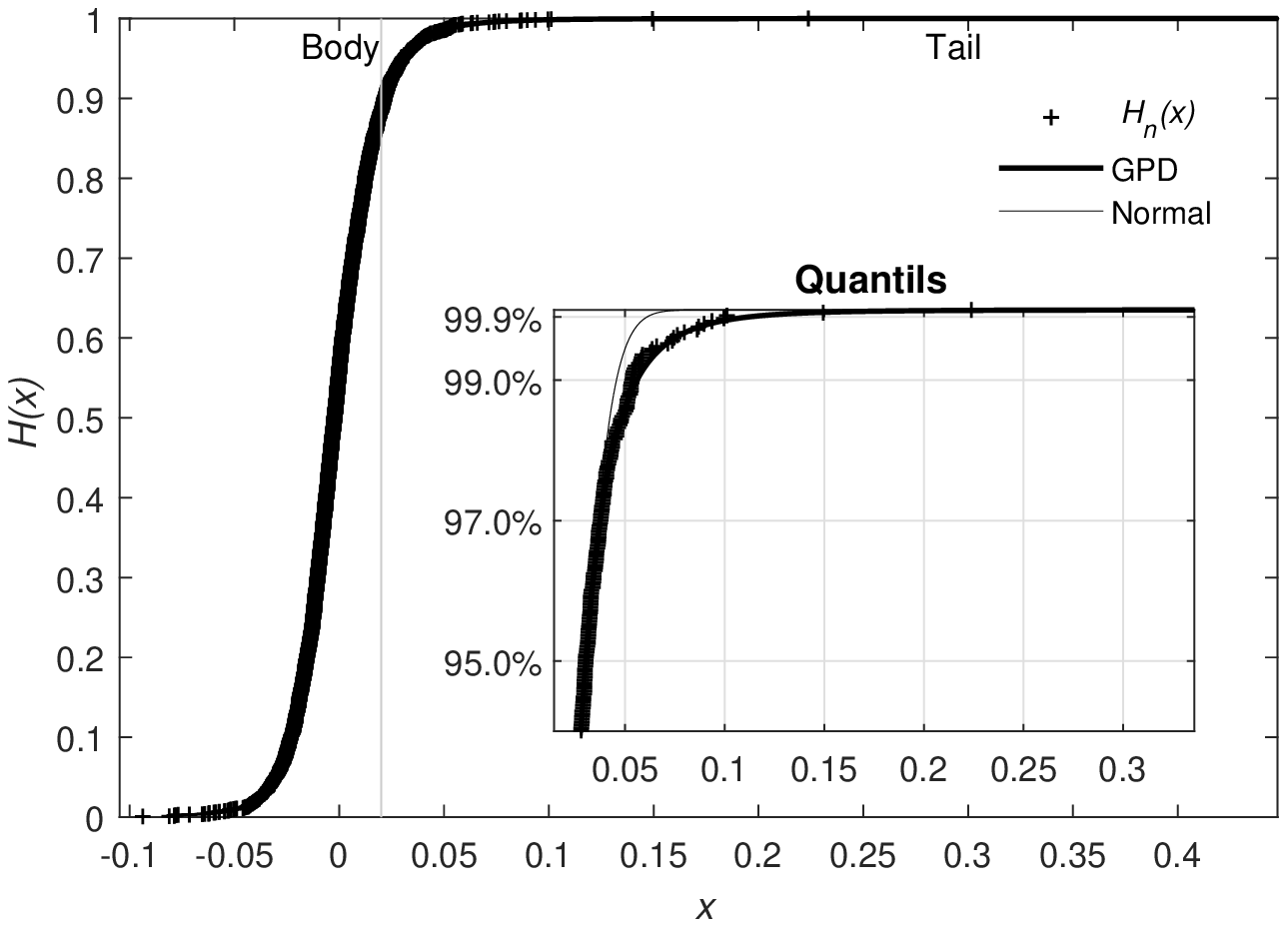}\\[-6pt]
	\begin{quote}
		\caption[Sample-Size 100]{\label{example2p2} The empirical density (upper figure) and the empirical distribution function (lower figure) for the weekly logarithmic rates of return. Additionally marked is the corresponding graph of the GPD as the tail model and the fitted normal distribution (inner picture of the lower figure).}
	\end{quote}
\end{figure}

Fig.\ \ref{example2p2} shows the density and distribution function of the adapted GPD in comparison to the empirical density and empirical distribution function.
At a threshold of $u = 0.020$, the body and the tail of the empirical density $h_n(x)$ and distribution $H_n(x)$ are separated. The inner picture of the lower figure shows the tail region and the fitted GPD (bold line) compared to the empirical distribution. These two graphs are superimposed, showing that any differences are difficult to distinguish.\\

If a geometric brownian motion is assumed for the weekly logarithmic returns of the MSCI World Index, then the parent distribution of the logarithmic returns should be a normal distribution.
The thin line in the lower figure of Fig.\ \ref{example2p2} shows the graph of a fitted normal distribution.
There are slight differences in the area of the tail. Whereas the graph of the GPD is hardly distinguishable from the data.
This suggests that the true (unknown) distribution in the area of the tails can be better described by the GPD.
\subsubsection{Monthly and annual sampling}\label{mscimonthly}
\begin{figure}[htbp]
	\centering
	\captionsetup{labelfont = bf}
	\includegraphics[width=0.975\textwidth]{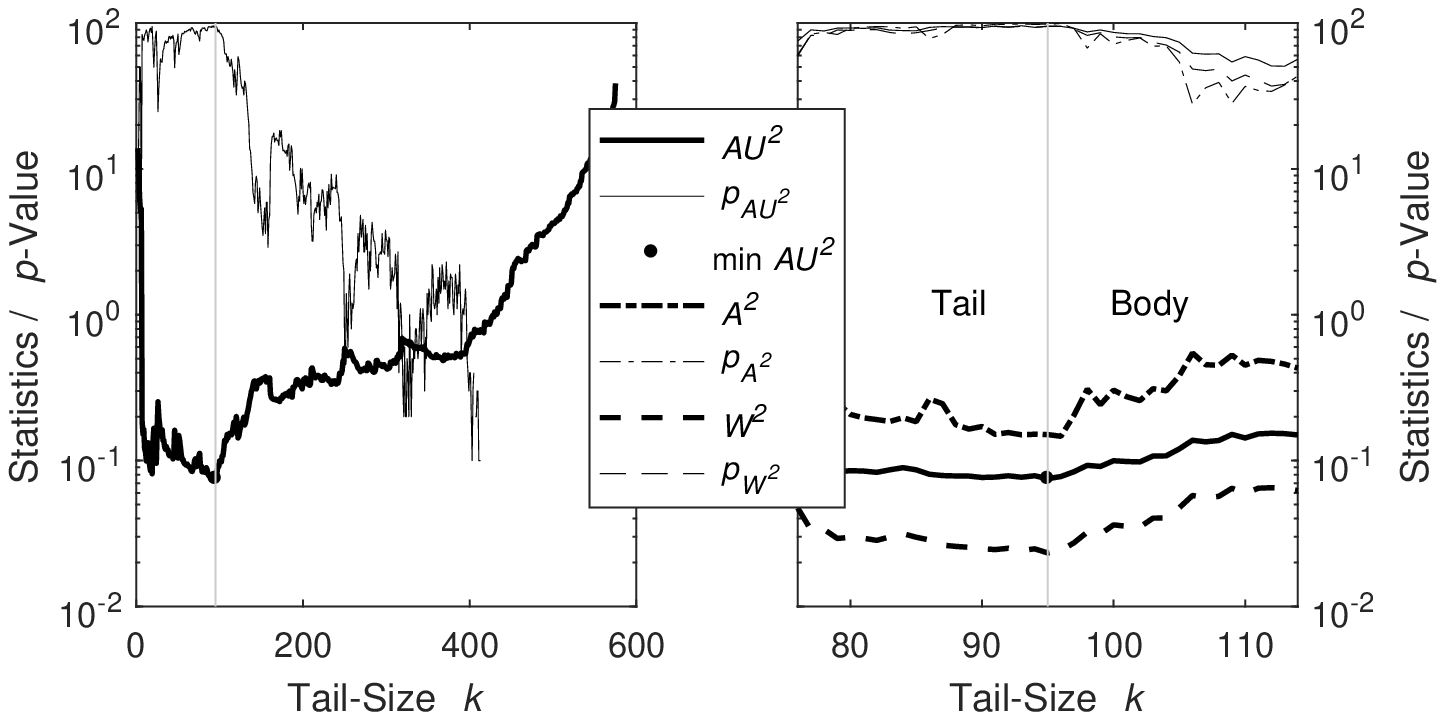}
	\includegraphics[width=0.975\textwidth]{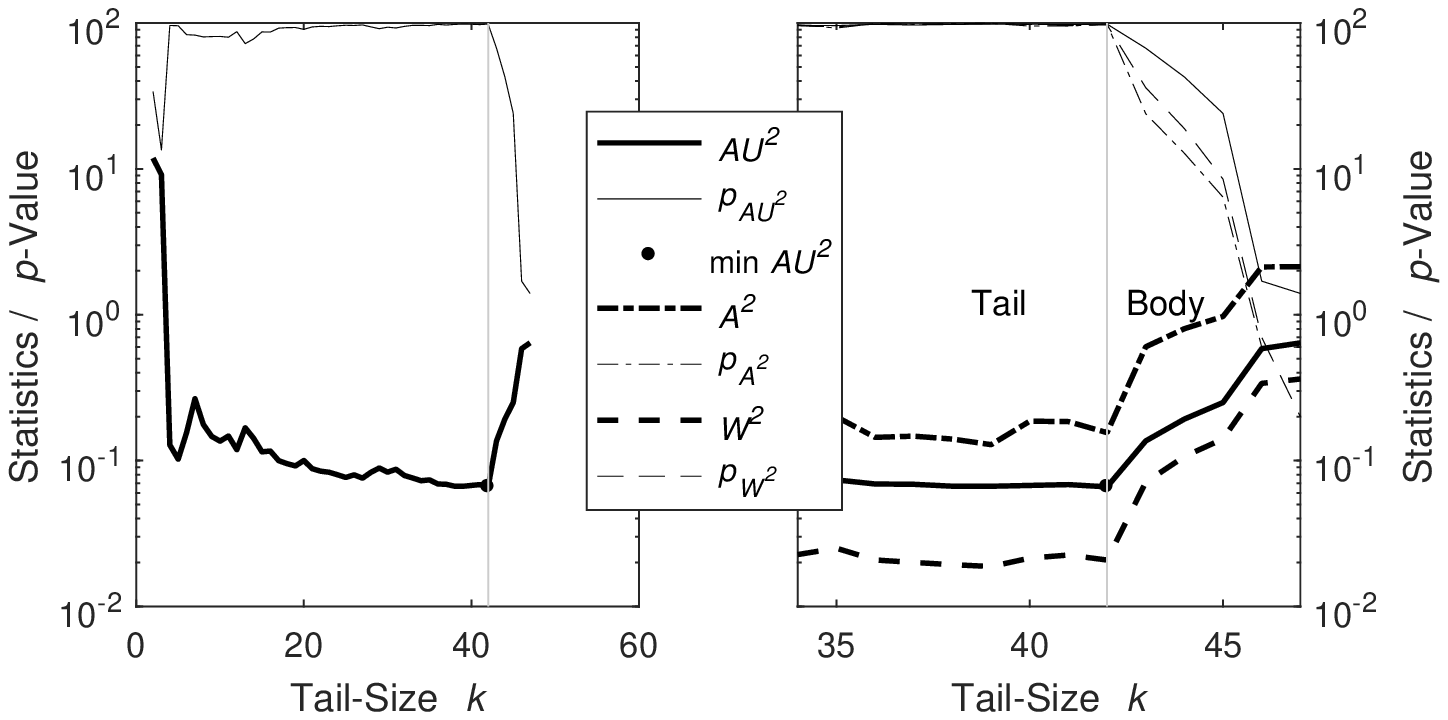}
	\begin{quote}
		\caption[Sample-Size 100]{\label{example2p3} Application of the procedure for detecting the beginning of the tail for $n=575$ monthly (upper figures) and $n=47$ annual (lower figures) logarithmic rates of return.}
	\end{quote}
\end{figure}

The same procedure is now applied to the monthly and yearly sampled data to find a suitable model for the tail of the unknown parent distribution. Fig.\ \ref{example2p3} shows, using the statistics and $p$-values, how the procedure works when these data are submitted to the program.
In the case of the monthly and annual values, we obtain $u = 0.027$ and $u = -0.216$, respectively, as thresholds, above which the adaptation of the GPD as a tail model seems favorable. 
This conclusion follows from the fact that for these thresholds, the statistics have small values and the $p$-values tend to be well above 95\%. 
Table \ref{example2results} summarizes these results and also shows the calculated parameters of the GPD. In addition, the risk indicators VaR and CVaR for the individual samples are listed in the lower part of the table.\\

The VaR corresponds to the quantile of the distribution for a given probability and can be calculated directly from the inverse function of the GPD given the distribution parameters. In forming the inverse function, however, the proportion of data to which the model refers for the tail of the unknown parent distribution must be considered. Note that not only in the ideal case examined in Section \ref{IDCA} but also in practice, as the sample length $n$ decreases, larger portions of the sorted time series are used to compute the model of the tail of the unknown parent distribution.
However, if the VaR is determined, the CVaR results by adding the value of the mean excess function of the GPD \citep{embrechts03}. Where in the calculation of the mean excess function, the location parameter is expressed by the VaR.\\

The risk indicators shown in the table refer to the logarithmic changes in the price $S$ of the MSCI World Index. To determine the risk capital that may need to be deposited, the results for the VaR and the CVaR can be recalculated in USD. For example, the VaR corresponds to a possible logarithmic (weekly, monthly or yearly) change in the MSCI World Index: $\text {VaR} = \log(S_1) - \log(S_0)$. A small algebraic transformation can be used to calculate the possible loss value in USD: $\Delta S = S_1 - S_0 = S_0 (\exp(\text{VaR})-1)$. This relationship follows directly from the property that the underlying process represents a geometric Brownian motion \citep{hull17}.
\begin{table}
	\captionsetup{labelsep=newline, justification=RaggedRight, singlelinecheck=false, labelfont = bf}
	\caption{Tail model and risk indicators for the MSCI World Index.}\label{example2results}
	\renewcommand{\arraystretch}{1.1}
	\begin{tabular}{lcrrr}
		\hline 
		&&&&\\[-6pt]
		                &                & \multicolumn{1}{l}{Weekly} & \multicolumn{1}{l}{Monthly} & \multicolumn{1}{l}{Annual}				\\[6pt]
Sample Size     		& $n$			 & 2503	  & 575     & 47 	   \\[6pt]
Threshold				& $k^*$			 & 289 	  & 95      & 42     \\
						& ${k^*}\!/{n}$  & 11.5\% & 16.5\%  & 89.4\% \\
						& $u$ 			 & 0.020		& 0.027	   & -0.216 		\\[6pt]	
GPD Parameter    		& $\hat{\xi}$    & 0.215        & -0.040         & -0.128   \\
						& $\hat{\sigma}$ & 0.011  	    &  0.037         &  0.205   \\[6pt] \hline
		&&&&\\[-6pt]
Statistic ($p$-Value)&$W_k^2$ & 0.021 (0.965) & 0.023 (0.955) & 0.021 (0.973)\\ 
		        	 &$A_k^2$ & 0.231 (0.845) & 0.151 (0.983) & 0.155 (0.984)\\
		      		 &$AU_k^2$& 0.078 (0.932) & 0.075 (0.948) & 0.066 (0.984)\\[6pt] \hline
		&&&&\\[-6pt]	
VaR (CVaR)			 & 95.0\%&0.030 (0.053) &0.070 (0.103) & 0.276 (0.427) \\
					 & 97.0\%&0.038 (0.062) &0.088 (0.121) & 0.347 (0.490) \\
					 & 99.0\%&0.056 (0.085) &0.126 (0.157) & 0.484 (0.611) \\
					 & 99.9\%&0.112 (0.157) &0.199 (0.227) & 0.715 (0.816) \\ [6pt] \hline
	\end{tabular}
\end{table} 
\section{Discussion and Conclusions}\label{CC}
With the procedure presented above, we can efficiently -- without any consideration of parameter specifications -- answer the following question: which threshold value divides an unknown parent distribution into the body and tail area? 
For this purpose, we have posed a theoretically sound method that, in addition to finding the optimal threshold value, determines the parameters of the GPD as a tail model based on the available data. Furthermore, the proposed method uses standard goodness of fit tests to show the quality of the fit.
In the examples shown, we have successfully tested the procedure on simulated and real data. However, limits to the application exist, as very small data sets ($n<25$) are difficult to analyze due to statistical fluctuations. For example, Fig.\ \ref{sample100} shows this phenomenon very well for a normal distribution, where the confidence interval (thin lines) contracts significantly only after the sample size exceeds 20.
Another limitation is caused by the assumption of the GPD as a tail model.
Although the distributions commonly used in the finance and insurance fields have the GPD as a tail model, exceptional parent distributions that do not have this property may exist. A further limitation could be the implementation costs.
From the current state of the research, it is difficult to estimate how far the proposed procedure will prevail. In individual cases, a business-related cost and benefit consideration will be the decision criterion for deployment. To facilitate the decision, we offer our procedure in a freely accessible Python software package.

The presented procedure offers numerous starting points for future research.
For example, it would be interesting to examine how the risk parameters determined for the single asset classes are presented in an aggregated portfolio and whether the overall risk can also be determined congruently from the historical portfolio data with the proposed procedure. 
Additionally, other research may consider the impact of our method on practice and regulatory affairs.

\end{document}